\documentclass[review]{elsarticle}

\usepackage{hyperref}
\usepackage{comment}
\usepackage{tabularx}
\usepackage{booktabs}

\journal{Information and Software Technology (IST)}

\begin{document}

\begin{frontmatter}

\title{Sentiment Analysis Tools in Software Engineering: A Systematic Mapping Study}

\author{Martin Obaidi\corref{mycorrespondingauthor}}  \author{Lukas Nagel} \author{Alexander Specht} \author{Jil Klünder}
\cortext[mycorrespondingauthor]{Corresponding author}
\address{Leibniz Universität Hannover\\ Software Engineering Group\\ Welfengarten 1\\ 30167 Hannover\\Germany}
\address{\{martin.obaidi, lukas.nagel, alexander.specht, jil.kluender\}@inf.uni-hannover.de}




\begin{abstract} 
\textbf{[Context]} Software development is a collaborative task. Previous research has shown social aspects within development teams to be highly relevant for the success of software projects. A team's mood has been proven to be particularly important. It is paramount for project managers to be aware of negative moods within their teams, as such awareness enables them to intervene. Sentiment analysis tools offer a way to determine the mood of a team based on textual communication.

\noindent \textbf{[Objective]} We aim to help developers or stakeholders in their choice of sentiment analysis tools for their specific purpose. Therefore, we conducted a systematic mapping study (SMS).

\noindent \textbf{[Method]} We present the results of our SMS of sentiment analysis tools developed for or applied in the context of software engineering (SE). Our results summarize insights from 106 papers with respect to (1) the application domain, (2) the purpose, (3) the used data sets, (4) the approaches for developing sentiment analysis tools, (5) the usage of already existing tools, and (6) the difficulties researchers face. We analyzed in more detail which tools and approaches perform how in terms of their performance.

\noindent \textbf{[Results]} According to our results, sentiment analysis is frequently applied to open-source software projects, and most approaches are neural networks or support-vector machines. The best performing approach in our analysis is neural networks and the best tool is BERT. Despite the frequent use of sentiment analysis in SE, there are open issues, e.g. regarding the identification of irony or sarcasm, pointing to future research directions.

\noindent \textbf{[Conclusion]} We conducted an SMS to gain an overview of the current state of sentiment analysis in order to help developers or stakeholders in this matter. Our results include interesting findings e.g. on the used tools and their difficulties. We present several suggestions on how to solve these identified problems.
 
\end{abstract}

\begin{keyword}
\texttt{Social Software Engineering, Sentiment Analysis, Machine Learning, Systematic Mapping Study}
\end{keyword}

\end{frontmatter}


\section{Introduction}
\label{sec:introduction}

Software projects are usually realized by teams of developers due to their growing complexity \cite{10.1145/203330.203345,300082}. Previous research on methods used in software development processes has found about 60\% of development teams to be distributed globally \cite{Kuhrmann.2018}. Digital communication tools like e-mail, Slack\footnote{\url{https://slack.com/}}, or JIRA\footnote{\url{https://www.atlassian.com/de/software/jira}} are especially important for teams working in such a distributed environment \cite{5196929, 10.1145/1882362.1882435}. Sentiments extracted from these digital communication tools have been found to affect the productivity, task synchronization and job satisfaction of the entire software development team \cite{Graziotin.2014,10.1002/smr.1673,10.1145/2441776.2441812, Guzman.2014}. Therefore, the analysis of sentiments is especially important for project managers, who need to be aware of negative moods within their teams in order to be able to intervene.

Detecting bad moods is a goal pursued by many researchers (cf. \cite{Novielli.,Calefato.2017,Ahmed.2017,Chen.2019,Islam.2017,Islam.2018b}). So-called sentiment analysis tools attempt to determine the mood of a team using text-based communication. There is a number of different sentiment analysis tools developed and applied in various different contexts \cite{Murgia.2014, 10.1145/2661685.2661689, Calefato.2018, Islam.2018}. In this paper, we strived to get an overview of the state-of-research on sentiment analysis in software engineering (SE). By providing this overview, we aim to help developers or stakeholders to choose suitable tools or approaches for their specific purpose as well as to show researchers the possible problems and solutions to fill identified research gaps. By this, we try to proceed on as high a level as possible and not to analyze too deeply and specifically, in order to give inexperienced readers a thus clearer overview.

Our previous work \cite{10.1145/3463274.3463328} contributed a systematic literature review (SLR) revealing a list of application scenarios, the used data sources and classification approaches as well as the problems encountered during the development of sentiment analysis tools. In this paper, we expand on our research by including recently published literature. We also analyze the amount of publications per year and the publication venues.  Furthermore, we analyze the performance of the approaches and the sentiment analysis tools proposed in recent years and investigate how well they are suited as data sources for the classification of sentiments present in text-based communication.

This article is an extension of the paper "Development and Application of Sentiment Analysis Tools in Software Engineering: A Systematic Literature Review" published in "Evaluation and Assessment in Software Engineering" in 2021 \cite{10.1145/3463274.3463328}. In this extended version, we focus more on presenting an overview of the field of research. Hence, in contrast to the conference paper~\cite{10.1145/3463274.3463328},  we consider this paper rather a systematic mapping study then a systematic literature review.

\textit{Outline:} The rest of the paper is structured as follows:
In Section \ref{sec:related}, we present related work. The design of the mapping study and the research methodology are explained in Section \ref{sec:review}. The results are presented in Section \ref{sec:results}. In Section \ref{sec:discussion}, we discuss our results, before concluding the paper in Section \ref{sec:conclusion}.

\section{Related Work}
\label{sec:related}

\subsection{Secondary studies of sentiment analysis in general domain}
Kumar and Jaiswal \cite{Kumar.2020} conducted an SLR with the goal of advancing the understanding of the feasibility, scope, and relevance of studies that apply soft computing techniques for sentiment analysis. They considered tools which used Twitter data and identified research gaps in the field. These gaps include an incessant need to enhance the performance of the sentiment classification tools and the usage of other data sets like Flickr\footnote{\url{https://www.flickr.com/}} or Tumblr\footnote{\url{https://www.tumblr.com/}}.
Abo et al. \cite{Mohamed.2019} conducted a systematic mapping study dealing with sentiment analysis for Arabic texts in social media. Devika et al. \cite{Devika.2016} looked at different approaches to sentiment analysis. Among other approaches such as support-vector machine (SVM), Naive Bayes classifier, they explained rule-based as well as lexicon-based methods.
Maitama et al. \cite{Maitama.2020} performed a systematic mapping study, which contains an examination of aspect-based sentiment analysis tools and an investigation of their approach, technique, diversity and demography.
Kastrati et al. \cite{app11093986} conducted a systematic mapping study regarding sentiment analysis of students' feedback. They identified 82 relevant studies and found a trend towards deep learning. They also highlighted the need for structured data sets and the focus on expressed emotion and its detection as an outcome.
Baragash and Aldowah \cite{Baragash_2021} also conducted a SMS in the field of higher education. They found 22 relevant papers. They have identified several application domains such as "course evaluation" or "teaching quality evaluation". In addition, their SMS showed as a result that sentiment analysis can help to improve the quality of teaching process and the performance of teachers.

However, all these SMSs and SLRs are not related to SE, and the data or tools are not designed for the domain of SE. Consequently, no information about areas or motivation to use the tools in the context of software development is offered. Besides, a need for SE-specific tools has already been identified by many studies \cite{Ahmed.2017, Calefato.2018, Chen.2019, Ding.2018, Islam.2018, Imtiaz.2018, StackEmo, F.Calefato.2015, Umer.2020, Werner.2018}.

\subsection{Secondary studies of sentiment analysis in SE domain}

Several authors analyzed the use of specific sentiment analysis tools in SE.
Zhang et al. \cite{9240704} compared sentiment analysis tools like Senti4SD \cite{Calefato.2018} and SentiCR \cite{Ahmed.2017} with each other. In addition, they described models based on the neural network BERT \cite{Devlin.2019}, which were trained with data related to SE such as GitHub\footnote{\url{https://github.com/}} or Stack Overflow\footnote{\url{https://stackoverflow.com/}}. In their replication study, Novielli et al. \cite{Novielli.replication2021} explained some sentiment analysis tools (e.g. Senti4SD \cite{Calefato.2018}) in great detail and described the underlying data.

Similarly, other papers compared sentiment analysis tools in their accuracy and described them in terms of their operation \cite{N.Novielli.2018,Novielli.}. Other papers mentioned some tools, too, but only briefly described them without going into details \cite{Biswas.2019,Chen.2019,Islam.2018c}.
In contrast to our work, the authors did not follow a systematic approach to consider the broad range of existing literature and tools, but rather focused on specific papers only. They did not go into detail about why they chose these tools or data and what tools are available.

Lin et al. conducted a systematic literature review on opinion mining in SE \cite{BinLin.2021}. Their scope was wide and included not only polarity detection and emotion detection, but also politeness detection and trust estimation. They focused on the data sets available, the performance comparison of the available tools, and the issues specific to tool adoption and customization. Their goal is to help researchers and developers adapt better tools.

In their SLR, Sánchez-Gordón and Colomo-Palacios \cite{SANCHEZGORDON201923} focused on the emotions of software developers. For this purpose, they examined 66 papers and identified, among other findings, the unreliability of sentiment analysis tools. As their focus was on developer-expressed emotions, they suggested other measures such as self-reported emotions or biometric sensors.

However, the focus of their work was not to provide an overview of sentiment analysis in the SE domain, e.g., which application scenarios they had, what the tools were used for, or which tool/machine learning algorithm perform how on average. Some papers compared and evaluated multiple tools on multiple data sets, but not all possible tools and data sets were included. Moreover, they do not provide a systematic overview of the application scenarios or potential problems in the whole area of sentiment analysis in SE.

\section{Mapping Study}

In our previous paper, we conducted an SLR until the end of December 2020 \cite{10.1145/3463274.3463328}. In this paper, we added two more research questions and conducted a renewed search for the search space from January 2021 to the end of October 2021. We found 27 new papers, which passed all filtering steps and the quality assessment. Moreover, we have renewed the review of the previous found papers and removed one paper as a result. Below we have added the new data to our previous paper. The reference to the initial data set can be found in our previous paper \cite{10.1145/3463274.3463328}. However, all aspects of our study remained the same for the new search (e.g., search string or database selection), with the exception of excluding the forward search, which was not considered necessary given the time period chosen.

\label{sec:review}
In order to gain an overview of the current state of sentiment analysis in the context of SE, we conducted a mapping study. In particular, we strive towards reaching the following goal formulated as proposed by Wohlin et al. \cite{Wohlin.2012}:\\

\noindent \fbox{%
\noindent \parbox{\dimexpr\linewidth-2\fboxsep-2\fboxrule}{%
\underline{\textbf{Research Goal: }}\\
\textit{Analyze} existing literature 
\textit{for the purpose of} identifying widely used sentiment analysis methods and tools
\textit{with respect to} different application scenarios in software engineering
\textit{from the point of view of} a researcher
\textit{in the context of} a mapping study.}}

\subsection{Research question}
\label{subsec:rq}
In order to achieve the research goal and to analyze the literature on sentiment analysis in software engineering from different viewpoints, we pose the following research questions:

\noindent \textbf{RQ 1: } \textit{What are the main application scenarios for sentiment analysis in the context of SE?} As a first step, we want to get an overview of the broad area of possible application scenarios in which sentiment analysis is used in the context of software projects.

\noindent \textbf{RQ 2: } \textit{For what purpose is sentiment analysis used in the investigated studies?} Next, we want to analyze the different reasons why sentiment analysis is performed.

\noindent \textbf{RQ 3: } \textit{What data is used as a basis for sentiment analysis?} We want to get an overview of the data used to train or evaluate the tools. This way, we investigate which data is suitable as a basis for sentiment analysis in development teams -- both as training and/or test data.

\noindent \textbf{RQ 4.1: } \textit{Which approaches are used when developing sentiment analysis tools?} 

\noindent \textbf{RQ 4.2: } \textit{How do these approaches perform on average regarding accuracy and F1 score?} 

With these two questions, we gain an overview of good practices in the development of sentiment analysis in SE.  

\noindent \textbf{RQ 5.1: } \textit{Which already developed tools are used during sentiment analysis?} 

\noindent \textbf{RQ 5.2: } \textit{How do these tools perform on average regarding accuracy and F1 score?} 

In comparison to RQ 4.1 and RQ 4.2, here we want to gain an overview of already developed sentiment analysis tools in SE.  

\noindent \textbf{RQ 6: } \textit{What are the difficulties of these approaches?} Last, we analyze the advantages and disadvantages of the used approaches and existing tools, problems and difficulties, etc. These insights point to future research directions that should be investigated to improve the applicability and the outcome of sentiment analysis tools.

\subsection{Method}
To provide an overview of the development and application of sentiment analysis in the context of SE, we conducted an SMS. Our approach is based on the research process proposed by Kitchenham et al. \cite{KITCHENHAM20097,KitchenhamBA.2007} as well as Petersen et al. \cite{Petersen.2008} and comprises five steps which we describe in the subsequent sections.

\subsubsection{Database selection}
\label{subsec:databases}
Not all relevant papers can be found in all scientific databases. To minimize the risk of missing papers that are only available in a limited number of databases, we included a total of five databases in our search. Our selection follows the selections of other SMSs or SLRs in the SE domain \cite{8812836, 8984351, 7929422, kosa.2016, GAROUSI2016106, Klunder.2019, 10.1145/3029387.3029392, 10.1145/3379177.3388907} and consists of:
Science Direct\footnote{\url{https://www.sciencedirect.com/}}, 
IEEE Xplore\footnote{\url{https://ieeexplore.ieee.org/}}, 
ACM Digital Library\footnote{\url{https://dl.acm.org/}}, 
Springer Link\footnote{\url{https://link.springer.com/}}, 
and Google Scholar\footnote{\url{https://scholar.google.com/}}. 
We conducted a comprehensive search as proposed by Petersen et al. \cite{Petersen.2008} in each of these databases in order to reduce biases.

\subsubsection{Search string definition}
Our search string uses terms of the fields of \textit{sentiment analysis} and \textit{SE}. We also included the terms ``software project'' and ``development team''. While the former represents the typical use case of sentiment analysis in SE, the later represents the object whose sentiments are studied. Additionally, we found synonyms for sentiment analysis like ``opinion mining'' in related work \cite{8389299,Liu.2012,Liu2012}. Ultimately, we obtained the following search string:\\

\noindent \textit{(``Sentiment analysis'' \textbf{OR} ``text analysis'' \textbf{OR} ``opinion mining'' \textbf{OR} ``emotion AI'')}\\
\textit{\textbf{AND}}\\
\textit{(``software engineering'' \textbf{OR} ``development team'' \textbf{OR} ``software development'' \textbf{OR} ``software project'')}\\

\noindent We adjusted the search string according to the specific syntax of the data bases.

\subsubsection{Inclusion and exclusion criteria}
Publications that could not be used to answer our research questions were eliminated during the review process. Table \ref{tab:inclusion} contains the inclusion and exclusion criteria we applied to make these decisions more objective.
Every publication we found was tested with our exclusion criteria. Should a publication pass this test, we applied our inclusion criteria. Any publication that matches at least one inclusion criterion in this second check was included in our systematic mapping study. When a relevant publication appeared multiple times, for example due to multiple versions of the publication existing, we chose to include the version published most recently.

\begin{table}[htb!]
  \caption{Inclusion and exclusion criteria}
  \label{tab:inclusion}
  \begin{tabularx}{\columnwidth}{lX}
    \toprule
   \multicolumn{2}{l}{Inclusion}\\
    \midrule
    1. &The publication presents an approach of the application of sentiment analysis in the context of SE.\\
    2. & The publication presents an approach to creating an sentiment analysis tool/algorithm in the context of SE.\\
    3. & The publication addresses the research questions of this SMS in its goals, hypothesis or applications.\\
    \toprule
    \multicolumn{2}{l}{Exclusion}\\
    \midrule
    1. &The publication is not written in English.\\
    2. &The publication use data for sentiment analysis written in non-English language.\\
    3. &The publication is not peer-reviewed.\\
    4.& The publication appears repeatedly. In this case, we only considered the latest version.\\
    5.& The publication is not accessible (respectively only accessible only by payment).\\
    6.& The publication has technical content without proven scientific relevance such as invitation papers, editorials, tutorials, keynotes, speeches, white papers, grey literature, dissertations, theses, technical reports, and books.\\
    7.& The publication is a document that is not a full paper or study such as presentations, web postings, web content, citations, brochures, pamphlets, newsletters, or extended abstracts.\\
    \bottomrule
  \end{tabularx}
\end{table}

The inclusion criteria (see Table \ref{tab:inclusion}) were applied based on the results of the search by one of the authors to the best of his knowledge. In the search string, explicit words were used that belong to the domain of SE and sentiment analysis. Therefore, when there was any doubt during the selection process based on the title, we applied a pessimistic procedure. That is, the respective paper was rather included in the next filter step than excluded (if necessary, it was excluded as part of the next filter step). In addition, parts of the content could already be seen as a preview in 3 of 5 search engines when searching for the title. If the search terms specified in the search string were contained there, these papers were also considered for the next filtering step.

To better clarify point 3 of the inclusion criteria (see Table \ref{tab:inclusion}), we mention three examples of publications that were narrowly excluded and included in terms of content.

Calefato et al. \cite{F.Calefato.2015} explored what influences might lead to a response being marked as accepted on Stack Overflow. Among other aspects, they investigated the affect of comments by also calculating the sentiment value. Because they determined the sentiment value for the affect value and the paper addresses the content of SE, we included this paper.

Fang et al. \cite{Fang.2015}, on the other hand, focused on sentiment analysis by examining product reviews from Amazon.com. Although the paper is about Big Data and sentiment analysis, the paper does not mainly address SE content, so we excluded this paper.

Ortu el al. \cite{Ortu.2016} provided a data set built from JIRA issue comments. They assigned emotions like joy or sadness to the comments based on Parrot's framework \cite{parrott2001emotions}. Although the SE domain is addressed and it is about the topic of sentiment analysis, we did not include this paper based on inclusion criteria 1-3. That is  because this paper only provides a data set and did not apply a specific sentiment analysis or develop a new method in the sense of the inclusion criteria.

\subsubsection{Quality assessments}
We defined quality assessments according to Kitchenham et al. \cite{KITCHENHAM20097,KitchenhamBA.2007}. This allowed us to assess the quality of papers in the most objective manner. One example of our quality assessments is whether a paper provides comprehensible conclusions. We scored publications depending on whether they fulfilled the criterion fully (2), partially (1) or not at all (0). Any criteria that was not applicable was not taken into account. We calculated an average of all applicable quality criteria to decide whether a paper was of sufficient quality. Five publications whose average score was less than 1 were removed from further analysis.

\subsubsection{Execution}
\begin{figure}[htb!]
  \centering
  \includegraphics[width=0.80\textwidth, keepaspectratio]{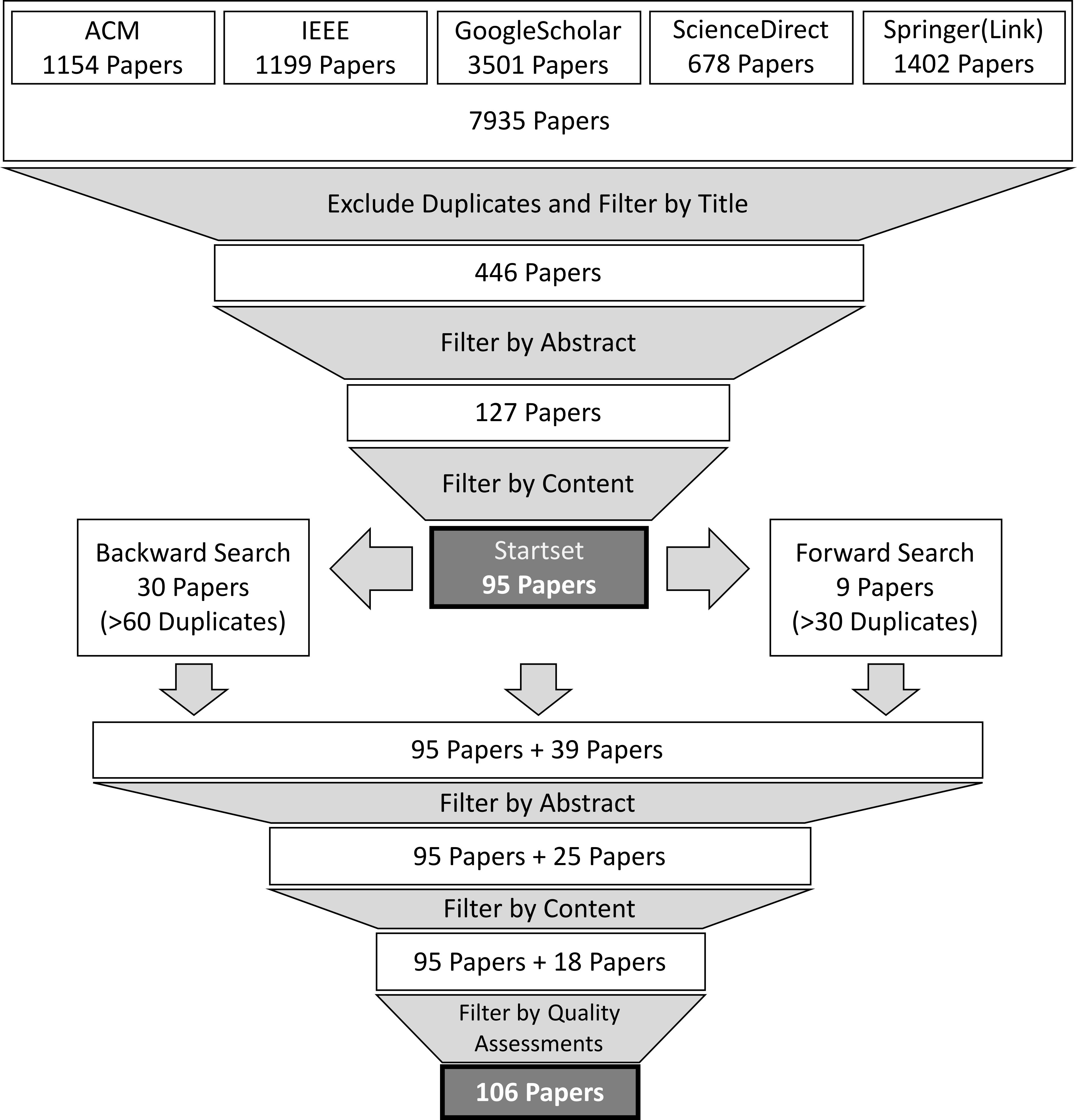}
  \caption{Search process and filtering steps}
  \label{fig:process}
\end{figure}

An overview of the execution can be seen in Figure \ref{fig:process}. The papers we considered for this work have been added to the numbers of our previous work in the Figure \ref{fig:process} accordingly. The initial numbers can be found in our previous paper \cite{10.1145/3463274.3463328}. 
We started off by searching the five databases mentioned in subsection \ref{subsec:databases} using the search string. From the 7935 results we selected potentially relevant publications based on their title. Once we removed duplicates, we applied the inclusion and exclusion criteria to the remaining 446 abstracts. We then scanned the full text of 127 papers, before using 95 papers as the startset for our forward and backward search, a snowball principle defined by Wohlin \cite{Wohlin.2014}. While the backward search (ca. 2500 including duplicates) resulted in 30 new papers that seemed relevant based on their title, the forward search (ca. 1000 including duplicates) yielded another 9 publications. The search engine Google Scholar\footnote{\url{https://scholar.google.com/}} was used for the forward search as mentioned by Wohlin \cite{Wohlin.2014}. Once again we scanned the 39 new papers, leading to 18 further inclusions in our results.

In total, we identified 113 relevant papers. We did not repeat the snowball principle since an examination of the 18 new papers found in the first iteration did not lead to any further relevant papers based on their titles. We ended our original search for publications in October of 2021.

Next, we applied the quality metrics to the 113 relevant papers. A total of seven publications were excluded due to their lack of research questions, a discussion of the results or a description of the technologies used. All seven excluded publications presented an average quality assessment score of less than 1. During this phase of our process, we also listed each relevant publication in an Excel spreadsheet [dataset] \cite{martin_obaidi_2021_4726650}, that consists of relevant parts of the paper. Based on this information, we clustered the papers and created categories related to our research questions.

\subsubsection{Validity procedures}
A number of factors influence the validity of our mapping study. For example, we cannot assume to have found all relevant publications. Similarly, we also cannot assume that our results are complete. It is likely that there are relevant publications that we could not include in our results, since we found a few papers that were not accessible.

In the following, we present threats our validity according to the different steps of the mapping study.

\textit{Database Selection}: The selection of databases performed in our SMS presents a threat to the construct validity. Using a large number of databases means that as many relevant papers as possible can be found. Some publications are listed in multiple databases, while others are only listed once. For our SMS, we searched five scientific databases that have been used by several other SMSs or SLRs in the SE domain \cite{8812836, 8984351, 7929422, kosa.2016, GAROUSI2016106, Klunder.2019, 10.1145/3029387.3029392, 10.1145/3379177.3388907}.

\textit{Search String Definition}: 
Another threat to the construct validity is our definition of the search string. We composed the search string using terms from the two areas \textit{SE} and \textit{sentiment analysis}. We also included synonyms for \textit{sentiment analysis}. Nevertheless, there may be other synonyms or related terms that we missed. We also did not use very specific SE terms in the search string to increase our search space and potentially find more papers. But, we cannot guarantee that we missed papers as a result of this choice (construct validity). However, we are confident that the accuracy of the results of our search string suffices to answer our research questions and therefore to achieve our goal of getting an overview of existing literature.

\textit{Inclusion and Exclusion Criteria}: Publications examined during our SMS were included or excluded based on various characteristics. For example, publications that has not been peer-reviewed were excluded. To improve on the internal validity of our process, we formulated our inclusion and exclusion criteria based on work by Kitchenham et al. \cite{KITCHENHAM20097,KitchenhamBA.2007}. Some criteria are purely objective, while others are still somewhat subjective. When in doubt, we included the paper and reviewed this decision at a later stage of our process.

\textit{Quality Assessments}:  We also set up quality assessments according to Kitchenham et al. \cite{KITCHENHAM20097,KitchenhamBA.2007}. Thus, a objective framework for the assessment of a publication's quality was provided. The framework allowed us to assess whether certain standards are met by papers. We chose the lower limit to exclude publications as a value of 1 which meant that all included publications fulfilled all applicable quality criteria at least partially on average.

\textit{Execution}: Most of the SMS was conducted by a single researcher. However, the whole process was reviewed by another. We expect the internal validity of our results to be impacted by some researcher bias, but we are also confident that this influence is only small.

\textit{Results}: We cannot guarantee that we found all relevant publications. Applying the snowball principle repeatedly could have lead to the discovery of other potentially relevant papers. However, we believe that the papers included in our SMS are sufficient to answer the research questions.

\section{Results}
\label{sec:results}
As visualized in Figure \ref{fig:process}, our systematic mapping study revealed 106 publications relevant to sentiment analysis in the SE domain published since 2012. The raw data set of these publications is available online [dataset] \cite{martin_obaidi_2021_4726650}.

In the following subsections, we present our results according to their demographics and the research questions described in Section \ref{subsec:rq}.

\subsection{Demographics}

There have been some publications in 2013 (5), but an increased interest is identifiable since 2017 (14). However, the most attention was gained within the past four years as can be seen in Table \ref{tab:years}.




\begin{table}[htb!]
\centering
\caption{Publication by year}
\label{tab:years}
\begin{tabular}{lllllllllll}
\toprule
& \rotatebox[origin=l]{90}{2012} & \rotatebox[origin=l]{90}{2013} & \rotatebox[origin=l]{90}{2014} & \rotatebox[origin=l]{90}{2015} & \rotatebox[origin=l]{90}{2016} & \rotatebox[origin=l]{90}{2017} & \rotatebox[origin=l]{90}{2018} & \rotatebox[origin=l]{90}{2019} & \rotatebox[origin=l]{90}{2020}& \rotatebox[origin=l]{90}{2021} \\ \midrule
Total       & 1 & 5 & 8 & 8 & 6 & 14 & 19 & 10  & 15 & 20 \\  \bottomrule  
\end{tabular}
\end{table}


The $n=106$ papers we found were published in a total of 63 different publication venues. An overview of these can be seen in Table \ref{tab:conferences}. For the sake of clarity and conciseness, we only listed publication venues that occurred at least three times. All other are summarized as “Other”.

\begin{table}[htb!]
\centering
\caption{List of Publication venues found in our SMS}
\label{tab:conferences}
\begin{tabular}{llllllllll}
\toprule
& \rotatebox[origin=l]{90}{MSR} & \rotatebox[origin=l]{90}{SEmotion} & \rotatebox[origin=l]{90}{EMSE} & \rotatebox[origin=l]{90}{ICSME} & \rotatebox[origin=l]{90}{ESEM} & \rotatebox[origin=l]{90}{ESEC/FSE} & \rotatebox[origin=l]{90}{JSS} & \rotatebox[origin=l]{90}{SANER} & \rotatebox[origin=l]{90}{Other} \\ \midrule
Total       & 12 & 9 & 7 & 5 & 4 & 3 & 3 & 3 & 60  \\  \bottomrule  
\end{tabular}
\end{table}

The Mining Software Repositories Conference (MSR) has the most papers with 12. This is followed by the International Workshop on Emotion Awareness in Software Engineering (SEmotion) with 9. 
The Journal Empirical Software Engineering (EMSE) follows in third place with 7 publications.
The aggregated publication venue "Other" includes e.g. the International Conference on Evaluation and Assessment in Software Engineering (EASE) or the International Conference on Software Engineering (ICSE) with 2.

The first author Md Rakibul Islam, together with the author Minhaz F. Zibran, has the most papers we found in our SMS with 8 (e.g. \cite{Islam.2016, Islam.2017}). Nicole Novielli was the most frequent author with 9 times in total (e.g. \cite{Novielli., Calefato.2017}).

\noindent \fbox{%
\noindent \parbox{\dimexpr\linewidth-2\fboxsep-2\fboxrule}{%
\underline{\textbf{Finding:}} Most publications in the context of sentiment analysis in SE were published from 2017 onwards, with a peak 2021 with 20. The Mining Software Repositories Conference has the highest number of publications with 12.}}

\subsection{Application domain}
In total, we identified three application domains for sentiment analysis: (1) open-source software (OSS) projects, (2) industry and (3) academia. Publications that did not name the application domain explicitly were manually classified according to the used data sets or the context of use. For example, if a publication indicated the use of GitHub, we assigned it to the OSS domain. Similarly, publications that used data sets like app reviews were assigned to the industry domain.

Of our total of $n=106$ papers, 79 were classified as belonging to the OSS domain, 24 to the industry domain and 7 to the academia domain. Paper in the OSS domain often analyzed open source data from platforms like GitHub \cite{Ding.2018, 10.1145/3424308, 9240704, 9492202} or Stack Overflow \cite{10.1145/3424308, Calefato.2019, 9240704, 9492202}. A sentiment analysis performed on the chat communication data of developers at Amazon MTurk \cite{10.1145/3392877} presents one example of an application in the industrial context. Academic applications include an analysis of communication among students in software projects organized by universities \cite{Gkontzis.2017,Guzman.2013b,Guzman.2013}.\\

\noindent \fbox{%
\noindent \parbox{\dimexpr\linewidth-2\fboxsep-2\fboxrule}{%
\underline{\textbf{Finding:}} Most of the papers on sentiment analysis in SE are based on open-source projects. Less then 1/3 of the papers either considers industrial projects or the academia.}}


%
    
\subsection{Purpose}
\label{sec:rq2}
The found papers can be split up in three types of papers based on their motivation: (1) development, (2) comparison, and (3) application of sentiment analysis tools. We see this categories as primary types that can be refined further.
Publications of the \textit{development} type aim to develop sentiment analysis tools. Other papers focus on the \textit{comparison} of already existing tools or their \textit{application}. Some papers presented a newly developed tool and compared it to existing alternatives in the same publication. In these cases we assigned them to the type \textit{development} as their primary goal was the development of a new tool.

The first type, \textit{development}, consists of papers that aim to develop a sentiment analysis tool. The second type compares already existing tools with each other, while the third type focuses on their application. There are papers that developed a new tool and then compared it to existing tools. Since all these papers had the primary goal of developing a new tool, we assigned them to the type \textit{development}.

Of our total of $n=106$ papers, we assigned 21 to the \textit{development} type, 14 focused on \textit{comparisons} and the remaining 71 dealt with the \textit{application} of sentiment analysis tools. Further details can be found in Table \ref{tab:rq2}. Papers assigned to the \textit{development} type focus on new procedures for sentiment analysis examining certain data in the SE context. One example is a sentiment analysis of developer communication \cite{Calefato.2018}. Eleven papers of the \textit{comparison} type attempt to find the best sentiment analysis tool. The other three compare the allocation of sentiments resulting from sentiment analysis tools with allocations given by humans. Of the $n=71$ papers of the \textit{application} type, 45 examine correlations between sentiments and specific values. Twenty-five analyze social aspects of developers. A total of 17 papers attempt to measure specific values related to sentiment analysis like the subjective usability \cite{ElHalees.2014} or the marketability of an open-source app \cite{Nayebi.2017}. Eight papers use sentiment analysis tools in combination with other methods to predict special values like the performance of a teacher based on feedback from students \cite{.2015,Aung.2017}.

\begin{table}[htbp!]
\centering
\caption{Motivations of the papers in detail}
\label{tab:rq2}
\begin{tabular}{lllllll}
\toprule
Type        & \rotatebox[origin=l]{90}{Find best tool} & \rotatebox[origin=l]{90}{Tools vs. human} & \rotatebox[origin=l]{90}{Correlations}  & \rotatebox[origin=l]{90}{Social aspects} & \rotatebox[origin=l]{90}{Values measurements} &  \rotatebox[origin=l]{90}{Values predictions} \\\midrule
Comparison  & 11    & 3      & 0              & 0           & 0       & 0\\
Application & 0    & 0     & 45              & 25           & 17       & 8\\
\bottomrule
\end{tabular}
\end{table}

\noindent \fbox{%
\noindent \parbox{\dimexpr\linewidth-2\fboxsep-2\fboxrule}{%
\underline{\textbf{Finding:}} There are three types of papers, based on their main purpose: Development of sentiment analysis tools, comparison and application of them. Most of the papers belong to the application type, whereas less than 33\% of the papers either belong to development or comparison.}}

\subsection{Used data sources}
We also analyzed the data used for the training and evaluation of sentiment analysis tools. We examined from which source (e.g. platform, website, etc.) the data was gathered. Thus, this does not mean that for example two papers assigned to the same data source used identical data sets. 

A total of 35 data sources was identified. The vast majority of these data sources were only used in one or two publications examined in our SMS. We believe this to be the case due to data sources in the application domain being unique in most cases. One example for such a unique data source is a set of software projects conducted in the context of university courses (e.g. \cite{Guzman.2013b,Guzman.2013}). For the sake of clarity and conciseness, we only listed data sets that occurred at least three times in Table \ref{tab:rq3}. All other data sets are summarized as ``Other''.

\begin{table}[htbp!]
\centering
\caption{Overview of the used data}
\label{tab:rq3}
\begin{tabular}{llllllll}
\toprule
Type        & \rotatebox[origin=l]{90}{Stack Overflow} & \rotatebox[origin=l]{90}{JIRA} & \rotatebox[origin=l]{90}{GitHub} & \rotatebox[origin=l]{90}{App reviews} & \rotatebox[origin=l]{90}{Code reviews} & \rotatebox[origin=l]{90}{Twitter} & \rotatebox[origin=l]{90}{Other}\\\midrule
Development & 12    & 8      & 4              & 2           & 2       & 2            & 2\\
Comparison  & 9    & 11      & 6              & 1           & 4       & 0            & 5\\
Application & 12    & 8     & 16              & 11           & 0       & 2            & 26\\ \bottomrule
Total       & 33   & 27     & 26             & 14           & 6       & 4            & 33\\
\bottomrule
\end{tabular}
\end{table}

Publications in the development and comparison categories use the same data source for training and testing purposes. One exception is a work by Chen et al. \cite{Chen.2019}, who use GitHub emojis to fine-tune an existing neural network, but perform their testing with another data set. The most frequently represented data sources in the publications examined in our SMS are \textit{Stack Overflow} (33 uses), \textit{JIRA} (27 uses), \textit{GitHub} (26 uses). App reviews were used as a data source in 14 papers and code reviews in 6 papers, while all others were only mentioned 4 times at most. As for the application domain, 22 out of 26 different data sources from "Other" were unique. Other data sources found in our SMS are chat data from Amazon MTurk \cite{10.1145/3392877}, android bug reports \cite{Umer.2020} and support tickets from IBM \cite{Werner.2018}.

\noindent \fbox{%
\noindent \parbox{\dimexpr\linewidth-2\fboxsep-2\fboxrule}{%
\underline{\textbf{Finding:}} There are a total of 35 different data sources. Six of them are used at least three times. Stack Overflow, JIRA and GitHub are used most frequently for training and testing sentiment analysis algorithms.}}

\subsection{Approaches for developing or using sentiment analysis}
\label{subsec:approaches}

In accordance with results presented in subsection \ref{sec:rq2}, we once again split all tools found in our SMS in the categories \textit{development} and \textit{application} depending on the context of the algorithm on which the sentiment analysis is based. For example, a paper dealing with the development of a new sentiment analysis tool which also gave information on the machine learning approach utilized by the approach, was assigned to the \textit{development} category. For publications making use of existing tools like SentiStrength \cite{Thelwall.2010,Thelwall.2012}, we listed the specific tools in Section \ref{subsec:tools}. There are also some papers that developed a new tool and compared it to existing ones within the same publication.

Table \ref{tab:rq4-1} provides an overview of the approaches being used during development. We did not list papers of the \textit{comparison} category, as those papers only compared existing tools. Therefore, our total number of publications here is $n=92$. For the sake of clarity and conciseness, we once again summed up all approaches that appeared less than three times as ``Other''.

\begin{table}[htb!]
\centering
\caption{Overview of approaches used during developing sentiment analysis tools}
\label{tab:rq4-1}
\begin{tabular}{llllllllll}
\toprule
Type    & \rotatebox[origin=l]{90}{Bayes} & \rotatebox[origin=l]{90}{Neural network} & \rotatebox[origin=l]{90}{SVM} & \rotatebox[origin=l]{90}{Lexicon/Heuristic} & \rotatebox[origin=l]{90}{Logistic regression} & \rotatebox[origin=l]{90}{Random forest} & \rotatebox[origin=l]{90}{Decision tree} & \rotatebox[origin=l]{90}{Gradient boosting} & \rotatebox[origin=l]{90}{Other}\\ \midrule
Development & 5     & 6   & 7              & 4             & 3                   & 3             & 2                 & 3 & 8\\ 
Application & 5     & 4   & 3              & 1             & 1                   & 1             & 1                 & 0 & 1\\ \bottomrule
Total       & 10    & 10  & 10              & 5             & 4                   & 4             & 3                 & 3  &  9\\ \bottomrule   
\end{tabular}
\end{table}

A total of 28 out of 92 publications provided information on which approaches were used. We found a total of 15 different machine learning approaches used for evaluation purposes in these papers. The most frequently used approaches were different kind of Bayes classifiers (e.g. Naive Bayes), neural network and SVM with 10 appearances. Other established methods like random forest, logistic regression or AdaBoost were also used. Five machine learning methods were only used once, namely bootstrap aggregating \cite{Ding.2018}, sequential minimum optimization \cite{Cagnoni.2020}, bootstrap aggregating \cite{Ding.2018}, pattern-based approach \cite{Lin.2019}, a voting classifier \cite{10.1007/978-3-030-64266-2_8} and word embedding \cite{10.1145/3442442.3458612}.

Eight papers compared different machine learning approaches to examine which worked best \cite{kumar2017opinion,Ahmed.2017,Ding.2018,Islam.2019,8890439,mostafa2018investigating,Murgia.2018,Patwardhan.2017}. The most frequent winner of these comparisons were approaches making use of SVMs (3 times). Gradient boosting won twice, while Bayes, logistic regression and neural network all won once.

We examined the machine learning methods in more detail for their average accuracy and max-averaged F1-score among all the $n=92$ papers, if these were given or could be calculated in the respective papers. An overview of this data can be seen in Table \ref{tab:rq4-2}. Because only existing tools were compared in the ``comparison'' category, we did not list the respective papers. For the sake of clarity, we only list all approaches that appeared more than three times regarding accuracy or F1 score. We also provided the number of values for the calculation of the average accuracy or F1 score. We have named this number "\# data points" in the Table. For example, the 15 data points in the average accuracy section of "Neural network" (see Table \ref{tab:rq4-2}) means that we found a total of 15 values in all $n=92$ papers that provided an accuracy of a neural network. The average of these 15 data points is then 0.87.

\begin{table}[htb!]
\centering
\caption{Overview of the average accuracy and F1 score of approaches used during developing sentiment analysis tools}
\label{tab:rq4-2}
\begin{tabular}{llllllllll}
\toprule
& \rotatebox[origin=l]{90}{Neural network} & \rotatebox[origin=l]{90}{SVM} & \rotatebox[origin=l]{90}{Adaptive boosting} & \rotatebox[origin=l]{90}{Random forest} & \rotatebox[origin=l]{90}{Gradient boosting} & \rotatebox[origin=l]{90}{Lexicon/Heuristic} & \rotatebox[origin=l]{90}{Decision tree} & \rotatebox[origin=l]{90}{Bayes} & \rotatebox[origin=l]{90}{Logistic regression}\\ \midrule
Average accuracy & 0.87  & 0.82 & 0.80   & 0.79           & 0.76        & 0.74          & 0.72     & 0.71 & 0.65\\
\# data points   & 15    & 8    & 1      & 6             & 2            & 7             & 2        & 4    & 2\\ \midrule
Average F1 score & 0,80 & 0,64  & 0,65   & 0,59           & 0,80        & 0,68          & 0,67     & 0,67 & 0,64 \\ 
\# data points   & 15    & 9   & 3       & 8             & 3            & 6             & 10       & 11    & 6\\ \bottomrule
\end{tabular}
\end{table}


From the 92 papers, 27 overall provided information about accuracy or F1 score of their used machine learning approach. Among these 27 papers, neural network had the best accuracy ($0.87$) and the best F1 score ($0.80$), averaged over 15 data points. Only gradient boosting was tied for the best F1 score with $0.80$, but only averaged over 3 data points. Decision tree and Bayes had over ten F1 scores, but their averages are both $0.67$. Compared to neural network with an average of $0.80$ over 15 data points, they both are significantly worse.  

\begin{figure}[htb!]
  \centering
  \includegraphics[width=0.70\textwidth, keepaspectratio]{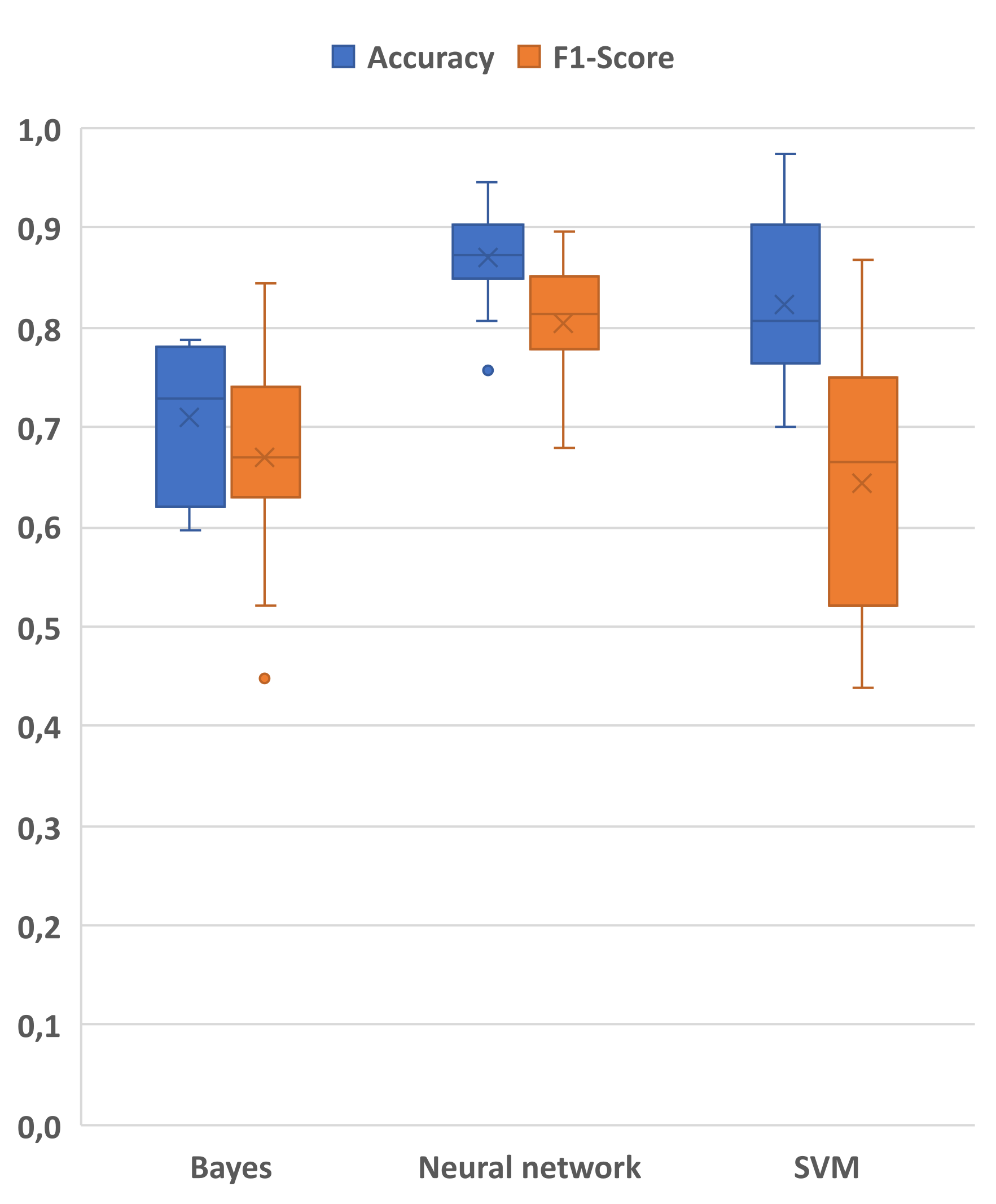}
  \caption{Boxplot of Bayes, neural network and SVM regarding accuracy and F1 score}
  \label{fig:boxplot-ml}
\end{figure}

Figure \ref{fig:boxplot-ml} shows the boxplot of the best performing approach among the most used approaches from Table \ref{tab:rq4-1}. 

\noindent \fbox{%
\noindent \parbox{\dimexpr\linewidth-2\fboxsep-2\fboxrule}{%
\underline{\textbf{Finding:}} Concerning the different machine learning methods there are 15, which were used for evaluation in the $n=92$ papers. Bayes, neural network and SVM stand out here. The authors often chose SVM because of its good performance. But regarding the average accuracy and F1 score among all the papers, neural network had the best performance. Only gradient boosting had also the best F1 score, but had 12 less data points than neural network.}}

\subsection{Existing tools for using sentiment analysis}
\label{subsec:tools}

Similar to \ref{subsec:approaches}, we investigate the usage of already existing tools and their performance. We found a total of 34 tools in the 106 publications considered in our SMS. For the sake of clarity and conciseness, we summed up all tools that appeared less than three times as ``Other''. In Table \ref{tab:rq5-1-2-approaches}, we list the tools (more than 2 appearances) and their used approach for applying sentiment analysis. 
\begin{table}[htb!]
\centering
\caption{Approaches of the tools. Tools marked with * are specifically designed for the SE domain)}
\label{tab:rq5-1-2-approaches}
\begin{tabular}{ll}
\toprule
Tools             & Approach \\ \midrule
SentiStrength     & Dictionary/Lexicon        \\
Senti4SD*         & SVM         \\
SentiStrength-SE* & Dictionary/Lexicon          \\
NLTK              & Dictionary/Lexicon         \\
SentiCR*          & Gradient Boosting Tree (GBT)        \\
CoreNLP           & Neural Network         \\
Vader             & Dictionary/Lexicon         \\
EmoTxt*           & SVM         \\
Alchemy           & NLP        \\
BERT              & Neural Network         \\
DEVA*             & Dictionary/Lexicon         \\
Syuzhet R         & Dictionary/Lexicon          \\
WNLU              & NLP         \\
WordNet           & Dictionary/Lexicon         \\ \bottomrule
\end{tabular}
\end{table}

For the tools WNLU and Alchemy we could not identify which approach was specifically used. We therefore used the general term NLP, which was also used in the description of the tools.

In Table \ref{tab:rq5-1}, we list the number of application of each tool. We also split all tools found in our SMS in the categories \textit{development} and \textit{application}. Furthermore, to distinguish whether the tools output emotion (E) or sentiment (S), we have provided their output in the table. It should be noted that specific emotions, which have been annotated based on an emotion model (e.g. Shaver et al. \cite{emotionlevels}), can be converted into sentiments (e.g. Calefato et al. \cite{Calefato.2018}). Moreover, tools such as BERT, which is a neural network, can also be given another output layer via transfer learning, so that they are more universally applicable and not limited to one output. Jithin et al. \cite{10.1145/3463274.3463805}, for example, used BERT \cite{Devlin.2019} to detect offensive language, but this could be interpreted as (negative) sentiment.



\begin{table}[htb!]
\centering
\caption{Overview of sentiment analysis tools applications. Tools marked with * are specifically designed for the SE domain)}
\label{tab:rq5-1}
\begin{tabular}{llllllllllllllll}
\toprule
Type & \rotatebox[origin=l]{90}{SentiStrength} & \rotatebox[origin=l]{90}{Senti4SD*} & \rotatebox[origin=l]{90}{SentiStrength-SE*} & \rotatebox[origin=l]{90}{NLTK} & \rotatebox[origin=l]{90}{SentiCR*} & \rotatebox[origin=l]{90}{CoreNLP} & \rotatebox[origin=l]{90}{Vader} & \rotatebox[origin=l]{90}{EmoTxt*} & \rotatebox[origin=l]{90}{Alchemy} & \rotatebox[origin=l]{90}{BERT} & \rotatebox[origin=l]{90}{DEVA*} & \rotatebox[origin=l]{90}{Syuzhet R} & \rotatebox[origin=l]{90}{WNLU} & \rotatebox[origin=l]{90}{WordNet} & \rotatebox[origin=l]{90}{Other}\\ \midrule
Output          & S             & S    & S                & S        & S        & S       & S     & E       & S      & S &  E/S &  S &  E/S &  S & - \\\midrule
Development     & 3             & 3    & 3                & 2        & 2        & 1       & 0     & 1       & 0      & 0 &  1 &  1 &  0 &  0 &  3\\
Comparison      & 9             & 6    & 6                & 7        & 6        & 5       & 1     & 1       & 3      & 3 &  2 &  0 &  1 &  1 &  8\\
Application     & 26            & 10    & 10              & 9        & 5        & 1       & 5     & 2       & 0      & 0 &  0 &  2 &  2 &  2 &  14\\ \bottomrule
Total           & 38            & 19   & 19               & 17       & 13       & 7       & 6     & 4       & 3      & 3 &  3 &  3 &  3 &  3 &  25\\ \bottomrule
\end{tabular}
\end{table}

Five tools stand out as being used at least 10 times. SentiStrength \cite{Thelwall.2010,Thelwall.2012} is the most used sentiment analysis tool by far with 38 uses. Its adaptation to the SE domain SentiStrenght-SE \cite{Islam.2017,Islam.2018} was used 19 times, together with Senti4SD \cite{Calefato.2018}. A natural languages toolkit handling sentiment analysis called NLTK \cite{Loper.17.05.2002} was used 17 times. SentiCR \cite{Ahmed.2017} was developed specifically for code reviews and was used 13 times. All other sentiment analysis tools were used less than 10 times, usually only 1 or two times in total. These include, but are not limited to, BERT-based model RoBERTa \cite{liu2019roberta}, WEKA \cite{396988} and the lexicon-based tool DEVA \cite{Islam.2018b}.\\

Similar to Section \ref{subsec:approaches}, we examined the sentiment analysis tools in more detail for their average accuracy and max-averaged F1-score among all the $n=106$ papers, if these were given or could be calculated in the respective papers. An overview of this data can be seen in Table \ref{tab:rq5-2}. Because only existing tools were compared in the ``comparison'' category, we did not list the respective papers. For the sake of clarity, we only list all approaches that appeared more than three times regarding accuracy or F1 score. We have also named the number of values considered in the average calculation as "\# data points".

\begin{table}[htb!]
\centering
\caption{Overview of the average accuracy and F1 score of sentiment analysis tools applications. Tools marked with * are specifically designed for the SE domain)}
\label{tab:rq5-2}
\small
\begin{tabular}{llllllllllllll}
\toprule
& \rotatebox[origin=l]{90}{BERT} & \rotatebox[origin=l]{90}{RoBERTa} & \rotatebox[origin=l]{90}{XLNet} & \rotatebox[origin=l]{90}{ESEM-E*} & \rotatebox[origin=l]{90}{ALBERT} & \rotatebox[origin=l]{90}{DEVA*} & \rotatebox[origin=l]{90}{SentiCR*} & \rotatebox[origin=l]{90}{EmoTxt*} & \rotatebox[origin=l]{90}{Senti4SD*} & \rotatebox[origin=l]{90}{SentiStrength-SE*} & \rotatebox[origin=l]{90}{SentiStrength} & \rotatebox[origin=l]{90}{CoreNLP} & \rotatebox[origin=l]{90}{NLTK} \\ \midrule
Average accuracy & 0.94      & 0.91    & 0.89     & 0.88   & 0.88     & 0.77  & 0.76  & 0.75 & 0.74   & 0.73  & 0.71   & 0.51   & 0.26\\
\# data points & 10        & 6       & 6        & 4      & 6        & 10    & 27    & 4    & 30     & 32    &  22    &  7     &  1 \\ \midrule
Average F1 score & 0.83      & 0.83    & 0.80     & 0.76   & 0.78     & 0.75  & 0.69  & 0.62 & 0.66   & 0.65  & 0.61   &  0.54  &  0.48 \\
\# data points & 12         & 6      & 6        & 4      & 6        & 10    & 36    & 9    & 44     & 44    & 32     &  15    &  9 \\ \bottomrule
\end{tabular}
\end{table}


From the 106 papers, 21 overall provided information about accuracy or F1 score of their used machine learning approach. The results show that BERT-based models like BERT \cite{Devlin.2019}, RoBERTa \cite{liu2019roberta} and ALBERT \cite{lan2020albert} performed very well, besides another neural network based tool XLNet \cite{NEURIPS2019_dc6a7e65}. The best tool in accuracy is BERT with $0.94$ among 10 data points. It also had the best F1 score with $0.83$, together with RoBERTa. It should be mentioned that ESEM-E \cite{Murgia.2018} and EmoTxt \cite{Calefato.2017} were explicitly designed to detect emotions such as fear or joy instead of polarities. The domain independent tools NLTK \cite{Loper.17.05.2002} and CoreNLP \cite{Manning.2014} have the worst accuracy and F1 values. Compared to BERT, they both have at least 43\% worse values in accuracy and 29\% worse values in F1 score.

\begin{figure}[htb!]
  \centering
  \includegraphics[width=0.80\textwidth, keepaspectratio]{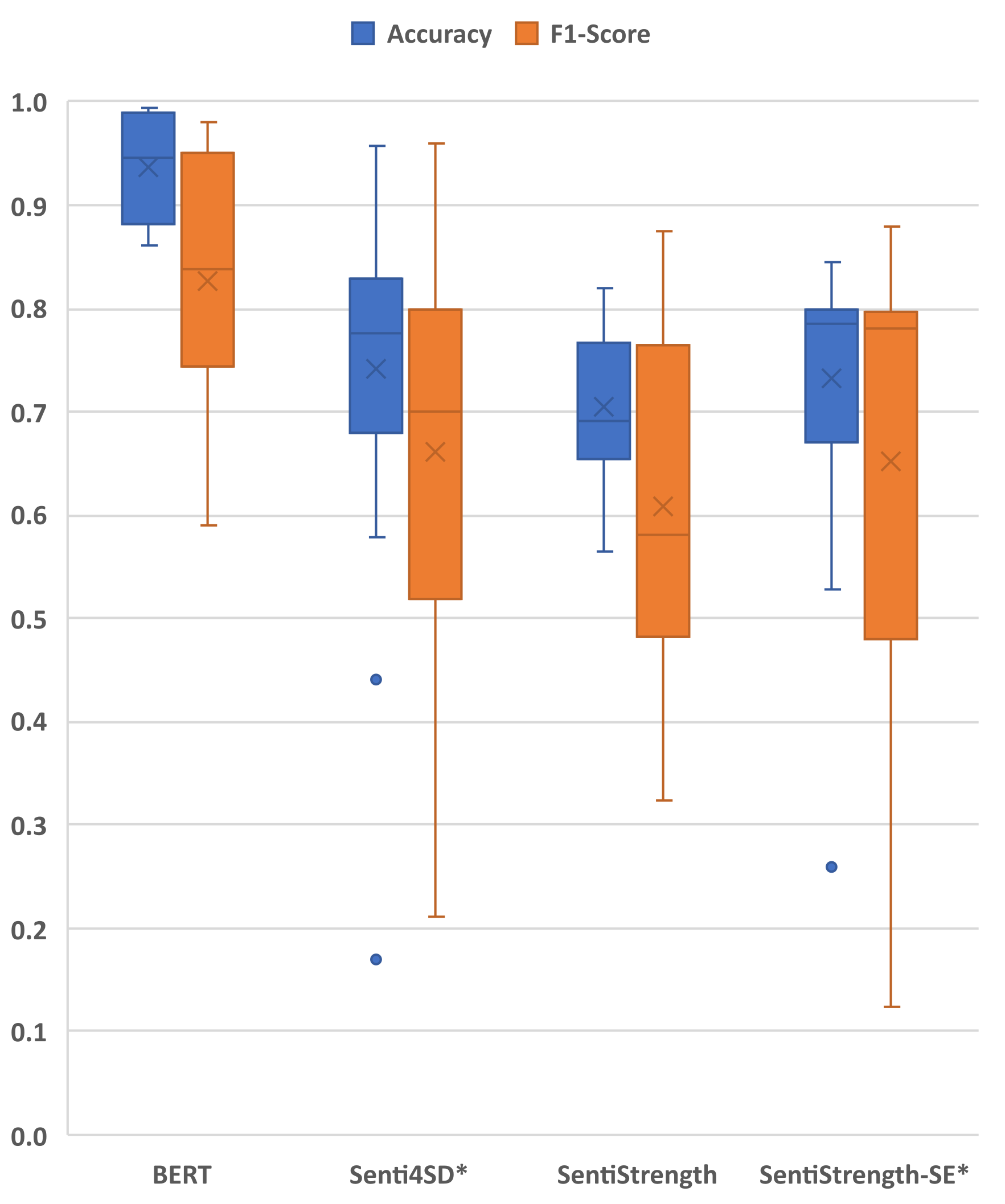}
  \caption{Boxplot of BERT, Senti4SD*, SentiStrength and SentiStrength-SE* regarding accuracy and F1 score. Tools marked with * are specifically designed for the SE domain}
  \label{fig:boxplot-tool}
\end{figure}

Figure \ref{fig:boxplot-tool} shows the boxplot of the best performing tool among the most used tools from Table \ref{tab:rq5-1}. BERT performs significantly better than the 3 most used tools Senti4SD, SentiStrength and SentiStrength-SE.

\noindent \fbox{%
\noindent \parbox{\dimexpr\linewidth-2\fboxsep-2\fboxrule}{%
\underline{\textbf{Finding:}} The results of our analysis show that 34 different existing sentiment analysis tools were used in the $n=106$ papers, with SentiStrength standing out. But regarding the overall performance, BERT-based tools had one of the best accuracies and F1 scores. The best tool is BERT with $0.94$ accuracy and $0.83$ F1 score, together with RoBERTa with also $0.83$.}}

\subsection{Difficulties}
\label{sec:difficulties}
A number of difficulties regarding the field of sentiment analysis in the SE domain were mentioned frequently among the examined papers. An overview of these issues is shown in table \ref{tab:rq6}.

\begin{table}[!htb]
\centering
\caption{Overview of mentioned problems regarding sentiment analysis in SE}
\label{tab:rq6}
\begin{tabular}{llllllllll}
\toprule
Type        & \rotatebox[origin=l]{90}{Adaption to the domain of SE} & \rotatebox[origin=l]{90}{Negation handling} & \rotatebox[origin=l]{90}{Irony/Sarcasm} & \rotatebox[origin=l]{90}{Subjectivity of manual labeling} & \rotatebox[origin=l]{90}{Cross-platform performance} & \rotatebox[origin=l]{90}{Small amount of data} & \rotatebox[origin=l]{90}{Disagreement}\\ \midrule
Development & 10           & 8           & 3               & 3             & 2                    & 6 & 0  \\ 
Comparison  & 6           & 3           & 4               & 5            & 4                    & 1  & 4 \\ 
Application & 13          & 3           & 6               & 5            & 2                    & 1 & 3 \\ \bottomrule
Total       & 29          & 14          & 13              & 13           & 8                    & 8   & 7 \\ \bottomrule
\end{tabular}
\end{table}

A total of 29 of the examined 106 papers mentioned the lack of adaptions of existing sentiment analysis tools to the SE domain \cite{Ahmed.2017,Calefato.2018,Chen.2019,Ding.2018,Imtiaz.2018}. Other problems like the handling of irony and sarcasm (13 times) or the subjectivity of a manual labeling of data (13 times) were also mentioned. Furthermore, there are also investigations on the performance of sentiment analysis tools when trained in cross-platform settings with 8 times. Novielli et al. \cite{Novielli.} found that tools trained with one data set usually performed poorly when tested with a different set of data. Several papers also found that there are disagreements among sentiment analysis tools \cite{Jongeling.2017, Kaur.2018, 10.1145/3479497, 8643972} or between tools and humans \cite{Jongeling.2017, Kaur.2018, 10.1145/3479497, 8643972, Imtiaz.2018, CABRERADIEGO2020105633, 9474673} regarding polarity or emotion perception.

\noindent \fbox{%
\noindent \parbox{\dimexpr\linewidth-2\fboxsep-2\fboxrule}{%
\underline{\textbf{Finding:}}
There a number of difficulties regarding sentiment analysis in the SE domain, which are mainly related to subjective data labels, too much customization for specific data sets and lacking adaptation to the SE domain.}}

\section{Discussion}
\label{sec:discussion}

\subsection{Answer to research questions}
\textit{RQ 1: What are the main application scenarios for sentiment analysis in the context of SE?}\\
We found three application domains: Open-source software domain, industry  and academia. More than 2/3 of all examined publications were classified in the OSS domain. One example of such a classification is the use of sentiment analysis on public data from OSS development platforms like GitHub or Stack Overflow \cite{10.1145/3424308, .2020, 9240704, 9492202}. However, we also found case studies applying sentiment analysis to chats within industrial \textbf{\cite{10.1145/3392877}} or scholarly \cite{Gkontzis.2017} developer teams.

~\\ \noindent
\textit{RQ 2: For what purpose is sentiment analysis used in the investigated studies?}\\
Publications found in our SMS could be split into three categories when looking at their purpose, namely development, comparison and application. A total of 71 of the $n=106$ papers focus on the application of a sentiment analysis tool. The authors were most often looking to find statistical correlations between sentiments and specific parameters. For example, one publication examined potential correlations of different times of a day and bug-introducing or bug-fixing commits \cite{.2018}. Furthermore, social aspects of developers were also studied in multiple works. One of these works by Whiting et al. \cite{10.1145/3392877} developed a new technique supporting online and remote teams to increase their viability by applying sentiment analysis to the developers' chats.

~\\ \noindent
\textit{RQ 3: What data is used as a basis for sentiment analysis?}\\
We found 35 different data sources, 6 of which were used more than twice. Stack Overflow (32 uses),JIRA (27 uses), GitHub (26 uses)  are the most frequently used data sources for training and testing purposes of new sentiment analysis algorithms. Most data sets found in our SMS consist of commit/pull request comments, discussions, questions, or reviews.

For the use of existing tools, we also found these three data sets being used. However, individual data sets from specific teams were most often used across all three application domains. This is also evident in our assignment of 92\% (23 out of 25) of all unique data sets found in papers of the application category. Some examples for these data sets originate from the industrial sector like chat data from Amazon's Mturk \cite{10.1145/3392877} or from software projects in universities \cite{Guzman.2013b,Guzman.2013}.

~\\ \noindent
\textit{RQ 4.1: Which approaches are used when developing sentiment analysis tools?}\\
We found a total of 15 different machine learning methods. The most used methods are Bayes, neural network and SVM with 10 uses. The results indicate that all these three approaches are the most frequently tested machine learning methods in the context of sentiment analysis tools. A commonly used tool that is not lexicon based and specifically designed for SE called Senti4SD \cite{Calefato.2018} implemented an SVM as it produced the best results. Publications comparing multiple machine learning methods also chose SVMs most often (3 times), while Gradient boosting was chosen second most (2 times).

~\\ \noindent
\textit{RQ 4.2: How do these approaches perform on average regarding accuracy and F1 score?}\\
Regarding the average accuracy and F1 score among all the approaches, neural network had the best performance with $0.87$ accuracy and $0.80$ F1 score, averaged over 15 data points in our analysis. Only gradient boosting had also $0.80$ F1 score, but it had 12 less data points than neural network in comparison. The second best approach in accuracy is SVM with $0.82$, but with $0.64$ F1 score.

~\\ \noindent
\textit{RQ 5.1: Which already developed tools are used during sentiment analysis?}\\
Our SMS yielded 34 different existing sentiment analysis tools. SentiStrength \cite{Thelwall.2010,Thelwall.2012} was used most often with 38 uses. We believe the reason for this frequent use to be the fact that the tool has been around for a while, is often been referred to as state-of-the-art, is domain-independent and does not require explicit training due to its use of a lexicon-based method.

~\\ \noindent
\textit{RQ 5.2: How do these tools perform on average regarding accuracy and F1 score?}\\
Regarding the overall performance of all tools, BERT stands out with the best accuracy ($0.94$) and F1 score ($0.83$) in our analysis, together with RoBERTa ($0.83$). Of the top five tools in accuracy and F1 score, four tools are based on neural networks.

~\\ \noindent
\textit{RQ 6: What are the difficulties of these approaches?}\\
One of the most prominent difficulties found in examined papers is that domain independent tools yield poor results in the SE domain \cite{Calefato.2018, Imtiaz.2018, 10.1145/3180155.3180195}. This is due to specific terms being used differently in the SE domain when compared to a non-technical context. This leads to different sentiments. Another difficulty is that the labeling of sentences with sentiments is often subjective, meaning that different people might assign different labels to the same sentences \cite{9240704, Imtiaz.2018, 10.1145/3180155.3180195,9462971}. Irony and sarcasm are also a problem mentioned in the examined publications, as a sentence has a different sentiment when it is known to be meant ironically \cite{Islam.2017, Islam.2018b, N.Novielli.2018}. It was also found that there are disagreements between tools and humans \cite{Jongeling.2017, Kaur.2018, 10.1145/3479497, 8643972, Imtiaz.2018, CABRERADIEGO2020105633, 9474673} or even between the tools themselves \cite{Jongeling.2017, Kaur.2018, 10.1145/3479497, 8643972}. Nevertheless, there is a temporal context to these problems, as some problems have already been addressed and are still being worked on.

\subsection{Interpretation}
The open availability of the data from OSS projects presents one possible explanation for the omnipresence of studies on this application scenario. This idea is further supported by most data sets used for training or evaluation belonging to the OSS domain. They are often extracted from platforms like GitHub and Stack Overflow. Public comments can be mined and labeled manually. However, to increase the performance of applications of sentiment analysis in the industry, tools should be trained with data taken from the industry domain. The importance of the use of domain specific data sets is further reinforced by the poor performance of existing tools in cross-platform settings.

The application category presented the tendency of too few application scenarios being done in the industry or that there is not enough interest for such tools in the industry context. Further investigative efforts might be necessary to find information on the demand for sentiment analysis in the SE domain and to find reasons for the lack of its use in relation to OSS. Legal and privacy issues might be one possible explanation for the scarcity of it use in the industry. Developers might have doubts whether their data is analyzed in terms of the adequacy of their language. This means that work councils often prohibit the analysis of existing data in terms of social aspects. Therefore, accessing data sets from the industry is much more complicated than using data available online which can be used with less restrictions.

SentiStrength \cite{Thelwall.2010,Thelwall.2012} is a well-established tool for sentiment analysis. It has been used frequently in the application category (26 times) and has also served as a comparison tool (9 times). It does not require any training as it is lexicon-based, rarely has performance drops in different domains and has a graphical user interface (GUI). The practical application of SentiStrength is often easier for most users than the application of machine learning based methods like BERT, where more knowledge of the respective programming language is required. However, such generic tools are not specialized for specific domains like SE. SentiStrength-SE \cite{Islam.2017} is an adaptation of the tool to the SE domain and was used second most commonly besides Senti4SD \cite{Calefato.2018}, possibly due to its adaptation to SE while still not requiring any training. Other tools from the $n=106$ papers examined in our SMS make use of machine learning approaches such as Bayes or SVM. These two machine learning techniques were the two most frequently used examples in the examined literature with 10 appearances, besides neural network. One tool using an SVM, Senti4SD \cite{Calefato.2018}, achieves better accuracy than lexicon-based tools such as SentiStrength-SE \cite{Islam.2017}, but performs significantly worse in cross-platform use cases \cite{Novielli.}. There are also other tools based on neural networks like RoBERTa \cite{liu2019roberta}, who outperform Senti4SD \cite{Calefato.2018} or other sentiment analysis tools \cite{9240704} in some cases. Our detailed analysis of the performance of the approaches and used tools indicated that the best performance was achieved with a neural network. Both the approaches and the tools themselves showed a clear tendency towards neural networks. Four out of five of the best performing tools in our analysis are based on a neural network. Three of them are Bert-based models. A conclusion from the frequency of usage to the performance is not possible without further research.

The publications examined in our SMS attempted to address some of the problems that are listed in subsection \ref{sec:difficulties}, such as the adaption of the SE domain. According to our results, the problem of irony or sarcasm is not yet solved. We found only two paper regarding this matter, which can be a first step towards solving this problem \cite{Fabry.2021, e23040394}. In addition, the existing tools usually perform differently in cross-platform use cases \cite{Novielli.,9240704,10.5555/2735522.2735528,Ahasanuzzaman.2020,9462971}. This means that a tool trained with a data set from GitHub will perform well on other data from GitHub, but worse on data collected from other platforms like Stack Overflow. Further investigations are necessary. Possible causes could be the choice of words being communicated differently across platforms, for example due to differing levels of politeness. Another possibility are the labels of different data sets. Some publications claim that people themselves often disagree on which polarity specific sentences have \cite{9240704,Imtiaz.2018,Kaur.2018,N.Novielli.2018,Novielli.,10.1145/3180155.3180195,Guzman.2017,Murgia.2018,Murgia.2014,Novielli.2015,S.Panichella.2015}. Furthermore, papers pointed out that tools often disagree on given data sets regarding sentiments, both among themselves \cite{Jongeling.2017, Kaur.2018, 10.1145/3479497, 8643972} and compared to human perception \cite{Jongeling.2017, Kaur.2018, 10.1145/3479497, 8643972, Imtiaz.2018, CABRERADIEGO2020105633, 9474673}. Mansoor et al. therefore focused on the different perceptions of developers and tools on sentiments in Stack Overflow questions and also found disagreement in a pilot study \cite{9474682}. This may indicate that the perception of polarities may be subjective and there is not much agreement between developers, unlike what was found by Murgia et al. \cite{Murgia.2014}. A more precise distinction regarding the perception of polarities can be useful. There can be different perception categories, for example when a developer reads a statement and either perceives sentiments in content (e.g. \textit{"This software is garbage"}), looks at the tone of the statement (e.g. \textit{"Not you again, Mike..."}), or tries to guess the context (e.g.\textit{"You're twisting my words a little bit, but it's ok :-)} or \textit{"Is this good to go then?"}). There may also be statements where you just cannot decide which polarity they might have. Therefore, it should also be investigated how confident the raters are in the selection of the respective polarity or emotion.

\subsection{Future research directions}
It could be useful to create data sets that emerge directly from the industry in order to address the low application of sentiment analysis in the industrial context. Sentiment analysis tools created on such data could perform better in the industry field. Nevertheless, subjective labeling and poor cross-platform performance are still present problems. Matching the collected communication data with additional regular sentiment survey data from developers could result in a mitigation of these issues. Such measures would enable comparisons between predictions of trained tools and manual sentiment data.

Training the tools requires good data. An overview of specific data sets not only in terms of their domain, but also their exact origin, number of data points (distribution of labels) and labeling process (ad-hoc vs. coding guide) would also be helpful to train the tools in a more precise and useful way to achieve the best possible noise-free results in multiple SE domains, as e.g. Novielli et al. \cite{N.Novielli.2018, Novielli.} had already observed. As a result, these tools could solve the problem of poor cross platform performance.

Examining data sets and their labels for subjectivity or labeling them according to a shared emotion model instead of ad-hoc annotations could prove to be another possible solutions for the poor cross-platform performance. Examples for these emotion models are the Plutchik model \cite{PLUTCHIK19803} or the PANAS scale \cite{Watson.1988}. A tool called \textit{BiasFinder} can detect bias among sentiment analysis tools, but focuses on characteristics like gender \cite{asyrofi2021biasfinder}. Furthermore, having the same author label different data sets would introduce the same bias to both sets of labels and therefore might make tools working on both data sets perform more uniformly. Ad-hoc annotations capture perceived sentiments more accurately than assignments based on emotion models. However, ad-hoc assignments are also more subjective. This could lead to inferior performances of machine learning techniques. Another possibility could be the combination of several well-performing tools. An investigation of the expressions and politeness levels used on different platforms and subdomains could also lead to better cross-platform results. For example, it should be investigated whether developers on GitHub communicate differently than on Stack Overflow.

One finding of the SMS is that the field of sentiment analysis seems to be more and more in demand according to our publication year analysis. Moreover, there are also many different data sets and tools used for training and evaluation. Therefore, a detailed analysis of all tools regarding all data sets would be useful.
In this context, a software can be created, which should be easy to use even for inexperienced users. The software should then have all tools implemented and enable the user to enter any data sets as input for training or evaluation. This allows users who want to use a certain tool or evaluate a specific data set based on our paper or similar papers to pursue their goal without in-depth expertise.

In cases where the data itself is of diminished quality, mining new data or even acquiring data from the industry domain could lead to better data sets.

The problem of irony handling is a domain independent one. New developments in natural language processing could be examined to find newly available approaches.

\subsection{Limitations and threats to validity}
In the following, we address the limitations and threats to validity in the context of our analysis and results.

Regarding research questions RQ1, RQ2, and RQ6, we classified the results into specific categories. These were established based on the results. However, there is a possibility that these categories are not representative or well chosen. However, the categories were reviewed by the other authors, so we are confident that they are appropriate.

Most of the categories we have established in the results have a clear distinction from each other. However, there may be borderline cases where the assignment may not be clear. Again, we are confident that based on the internal reviews, we have performed the assignment appropriately.

Since we do not distinguish which tools were evaluated on which data sets, we cannot be certain that the average performance we calculated is the same as the actual real performance when applying on new data. This also means that we cannot say with a high degree of certainty that the best approaches and tools found are always the best performing ones in relation to specific data sets. When calculating the average accuracy and F1 score, we do not distinguish whether or not the respective tool has been pre-trained on the respective data set, if possible. Many papers in the "Comparison" category have already covered this for various tools and data sets. Our goal was to provide a general performance indication that is calculated across all data sets and is not weighted or influenced by specific data sets with their respective distributions.
A different distribution (e.g., a data set with 90\% positive and 10\% negative statements) may lead to different results in the evaluation. But in some cases authors have applied measures against this in their training process. Furthermore, Novielli et al. \cite{Novielli.} have shown that the performance of different machine learning tools is significantly worse when they are pre-trained on e.g. GitHub data and evaluated on Stack Overflow data than when they are pre-trained on it as well. Both data sets have a very similar distribution and were annotated using the same emotion model from Shaver et al. \cite{emotionlevels}. Moreover, we are not able to decide which distribution in each scenario represents the real world, which is significant for the practical application. For example, if a data set from GitHub is built and annotated exactly equally distributed, but in reality on the platform is communicated much more positive than negative (e.g. 90\% positive statements are used in communication there), the tool pre-trained on the data set might not perform as well as a comparable tool pre-trained with a 90:10 distribution.

When calculating and analyzing the Average accuracy and F1 Score, we cannot exclude human errors. To minimize this threat, the calculations were additionally checked for correctness.

\section{Conclusion}
\label{sec:conclusion}
The social component of a developer plays a major role in development teams as there is a lot of communication via various channels. Therefore, appropriate interaction between developers using these channels is of great importance. Sentiment analysis can help mitigate negative moods affecting the productivity of development teams. To the best of our knowledge, no SMS has been conducted that provides an overview of the various tools, their development, application and problems that are described in literature regarding sentiment analysis regarding SE. We conducted such an SMS.

A total of 7935 papers were analyzed in total resulting in 106 relevant papers. We examined these publications in terms of their publication year, publication venue, application area of sentiment analysis tools, the underlying data, procedures and the purpose of application. One important finding is that most publications only utilize sentiment analysis instead of developing new tools or comparing tools' performances. Another finding is that based on results, the choice between tools or approaches should be neural networks based approaches like BERT, because they perform very well regarding accuracy and F1 score among all tools and approaches we investigated. 

We also identified several issues like the handling of irony and sarcasm or the limited amount of data available for the training and evaluation of tools based on machine learning. Other problems are the subjectivity of labeling data, the degradation of the performance in cross-platform settings and disagreements between tools themselves and humans. Our SMS resulted in some possible causes and solutions to these issues. According to our results, future research should focus on objectively labeled data and data from the industry. A deep analysis of existing data sets and, based on that analysis, a specific selection and implemented improvements could lead to better, noise-free data sets and thus to more precise training of tools. Furthermore, the already existing data sets should be examined for conflicting labels. Combinations of tools could also be considered in order to achieve better performance in cross-platform scenarios.

\section*{Acknowledgment}
This research was funded by the Leibniz University Hannover as a Leibniz Young Investigator Grant (Project \textit{ComContA}, Project Number \textit{85430128}, 2020--2022).


\bibliography{mybibfile}

\begin{thebibliography}{100}
\expandafter\ifx\csname url\endcsname\relax
  \def\url#1{\texttt{#1}}\fi
\expandafter\ifx\csname urlprefix\endcsname\relax\def\urlprefix{URL }\fi
\expandafter\ifx\csname href\endcsname\relax
  \def\href#1#2{#2} \def\path#1{#1}\fi

\bibitem{10.1145/203330.203345}
R.~E. Kraut, L.~A. Streeter, Coordination in software development, Commun. ACM
  38~(3) (1995) 69–81.
\newblock \href {http://dx.doi.org/10.1145/203330.203345}
  {\path{doi:10.1145/203330.203345}}.

\bibitem{300082}
D.~E. {Perry}, N.~A. {Staudenmayer}, L.~G. {Votta}, People, organizations, and
  process improvement, IEEE Software 11~(4) (1994) 36--45.
\newblock \href {http://dx.doi.org/10.1109/52.300082}
  {\path{doi:10.1109/52.300082}}.

\bibitem{Kuhrmann.2018}
M.~Kuhrmann, P.~Tell, J.~Kl{\"u}nder, R.~Hebig, {Sherlock A. Licorish}, S.~G.
  MacDonell, Helena stage 2 results (2018).
\newblock \href {http://dx.doi.org/10.13140/RG.2.2.14807.52649}
  {\path{doi:10.13140/RG.2.2.14807.52649}}.

\bibitem{5196929}
T.~Niinimaki, A.~Piri, C.~Lassenius, Factors affecting audio and text-based
  communication media choice in global software development projects, in: 2009
  Fourth IEEE International Conference on Global Software Engineering, 2009,
  pp. 153--162.
\newblock \href {http://dx.doi.org/10.1109/ICGSE.2009.23}
  {\path{doi:10.1109/ICGSE.2009.23}}.

\bibitem{10.1145/1882362.1882435}
M.-A. Storey, C.~Treude, A.~van Deursen, L.-T. Cheng, The impact of social
  media on software engineering practices and tools, in: Proceedings of the
  FSE/SDP Workshop on Future of Software Engineering Research, FoSER '10,
  Association for Computing Machinery, New York, NY, USA, 2010, p. 359–364.
\newblock \href {http://dx.doi.org/10.1145/1882362.1882435}
  {\path{doi:10.1145/1882362.1882435}}.

\bibitem{Graziotin.2014}
D.~Graziotin, X.~Wang, P.~Abrahamsson, Happy software developers solve problems
  better: psychological measurements in empirical software engineering, PeerJ 2
  (2014) e289.
\newblock \href {http://dx.doi.org/10.7717/peerj.289}
  {\path{doi:10.7717/peerj.289}}.

\bibitem{10.1002/smr.1673}
D.~Graziotin, X.~Wang, P.~Abrahamsson, Do feelings matter? on the correlation
  of affects and the self-assessed productivity in software engineering,
  Journal of Software: Evolution and Process 27~(7) (2015) 467--487.
\newblock \href {http://dx.doi.org/10.1002/smr.1673}
  {\path{doi:10.1002/smr.1673}}.

\bibitem{10.1145/2441776.2441812}
M.~De~Choudhury, S.~Counts, Understanding Affect in the Workplace via Social
  Media, Association for Computing Machinery, New York, NY, USA, 2013, p.
  303–316.

\bibitem{Guzman.2014}
E.~Guzman, D.~Az{\'o}car, Y.~Li, Sentiment analysis of commit comments in
  github: an empirical study, in: S.~Kim, M.~Pinzger, P.~Devanbu (Eds.), 11th
  Working Conference on Mining Software Repositories : proceedings : May 31 -
  June 1, 2014, Hyderabad, India, ACM, [Place of publication not identified],
  2014, pp. 352--355.
\newblock \href {http://dx.doi.org/10.1145/2597073.2597118}
  {\path{doi:10.1145/2597073.2597118}}.

\bibitem{Novielli.}
N.~Novielli, F.~Calefato, D.~Dongiovanni, D.~Girardi, F.~Lanubile, Can we use
  se-specific sentiment analysis tools in a cross-platform setting?,
  Proceedings of the 17th International Conference on Mining Software
  Repositories\href {http://dx.doi.org/10.1145/3379597.3387446}
  {\path{doi:10.1145/3379597.3387446}}.

\bibitem{Calefato.2017}
F.~Calefato, F.~Lanubile, N.~Novielli, Emotxt: A toolkit for emotion
  recognition from text, in: 2017 Seventh International Conference on Affective
  Computing and Intelligent Interaction Workshops and Demos (ACIIW), IEEE,
  Piscataway, NJ, 2017, pp. 79--80.
\newblock \href {http://dx.doi.org/10.1109/ACIIW.2017.8272591}
  {\path{doi:10.1109/ACIIW.2017.8272591}}.

\bibitem{Ahmed.2017}
T.~Ahmed, A.~Bosu, A.~Iqbal, S.~Rahimi, Senticr: A customized sentiment
  analysis tool for code review interactions, in: 2017 32nd IEEE/ACM
  International Conference on Automated Software Engineering (ASE), IEEE, 2017,
  pp. 106--111.
\newblock \href {http://dx.doi.org/10.1109/ASE.2017.8115623}
  {\path{doi:10.1109/ASE.2017.8115623}}.

\bibitem{Chen.2019}
Z.~Chen, Y.~Cao, X.~Lu, Q.~Mei, X.~Liu, Sentimoji: An emoji-powered learning
  approach for sentiment analysis in software engineering, in: Proceedings of
  the 2019 27th ACM Joint Meeting on European Software Engineering Conference
  and Symposium on the Foundations of Software Engineering, ESEC/FSE 2019,
  Association for Computing Machinery, New York, NY, USA, 2019, p. 841–852.
\newblock \href {http://dx.doi.org/10.1145/3338906.3338977}
  {\path{doi:10.1145/3338906.3338977}}.

\bibitem{Islam.2017}
M.~R. Islam, M.~F. Zibran, Leveraging automated sentiment analysis in software
  engineering, in: 2017 IEEE/ACM 14th International Conference on Mining
  Software Repositories, IEEE, Piscataway, NJ, 2017.
\newblock \href {http://dx.doi.org/10.1109/msr.2017.9}
  {\path{doi:10.1109/msr.2017.9}}.

\bibitem{Islam.2018b}
M.~R. Islam, M.~F. Zibran, Deva: sensing emotions in the valence arousal space
  in software engineering text, in: H.~M. Haddad, R.~L. Wainwright, R.~Chbeir
  (Eds.), Applied computing 2018, {Association for Computing Machinery Inc.
  (ACM)}, New York, NY, 2018, pp. 1536--1543.
\newblock \href {http://dx.doi.org/10.1145/3167132.3167296}
  {\path{doi:10.1145/3167132.3167296}}.

\bibitem{Murgia.2014}
A.~Murgia, P.~Tourani, B.~Adams, M.~Ortu, Do developers feel emotions? an
  exploratory analysis of emotions in software artifacts, in: S.~Kim,
  M.~Pinzger, P.~Devanbu (Eds.), 11th Working Conference on Mining Software
  Repositories : proceedings : May 31 - June 1, 2014, Hyderabad, India, ACM,
  [Place of publication not identified], 2014.
\newblock \href {http://dx.doi.org/10.1145/2597073.2597086}
  {\path{doi:10.1145/2597073.2597086}}.

\bibitem{10.1145/2661685.2661689}
N.~Novielli, F.~Calefato, F.~Lanubile, Towards discovering the role of emotions
  in stack overflow, in: Proceedings of the 6th International Workshop on
  Social Software Engineering, SSE 2014, Association for Computing Machinery,
  New York, NY, USA, 2014, p. 33–36.

\bibitem{Calefato.2018}
F.~Calefato, F.~Lanubile, F.~Maiorano, N.~Novielli, Sentiment polarity
  detection for software development, Empirical Software Engineering 23~(3)
  (2018) 1352--1382.
\newblock \href {http://dx.doi.org/10.1007/s10664-017-9546-9}
  {\path{doi:10.1007/s10664-017-9546-9}}.

\bibitem{Islam.2018}
M.~R. Islam, M.~F. Zibran, Sentistrength-se: Exploiting domain specificity for
  improved sentiment analysis in software engineering text, Journal of Systems
  and Software 145 (2018) 125--146.
\newblock \href {http://dx.doi.org/10.1016/j.jss.2018.08.030}
  {\path{doi:10.1016/j.jss.2018.08.030}}.

\bibitem{10.1145/3463274.3463328}
M.~Obaidi, J.~Kl\"{u}nder, Development and application of sentiment analysis
  tools in software engineering: A systematic literature review, in: Evaluation
  and Assessment in Software Engineering, EASE 2021, Association for Computing
  Machinery, New York, NY, USA, 2021, p. 80–89.
\newblock \href {http://dx.doi.org/10.1145/3463274.3463328}
  {\path{doi:10.1145/3463274.3463328}}.

\bibitem{Kumar.2020}
A.~Kumar, A.~Jaiswal, Systematic literature review of sentiment analysis on
  twitter using soft computing techniques, Concurrency and Computation:
  Practice and Experience 32~(1) (2020) e5107, e5107 CPE-18-1167.R1.
\newblock \href {http://dx.doi.org/10.1002/cpe.5107}
  {\path{doi:10.1002/cpe.5107}}.

\bibitem{Mohamed.2019}
M.~E.~M. Abo, R.~G. Raj, A.~Qazi, A.~Zakari, Sentiment analysis for arabic in
  social media network: A systematic mapping study (2019).

\bibitem{Devika.2016}
M.~Devika, C.~Sunitha, A.~Ganesh, Sentiment analysis: A comparative study on
  different approaches, Procedia Computer Science 87 (2016) 44 -- 49, fourth
  International Conference on Recent Trends in Computer Science \& Engineering
  (ICRTCSE 2016).
\newblock \href {http://dx.doi.org/10.1016/j.procs.2016.05.124}
  {\path{doi:10.1016/j.procs.2016.05.124}}.

\bibitem{Maitama.2020}
J.~Z. Maitama, N.~Idris, A.~Zakari, A systematic mapping study of the empirical
  explicit aspect extractions in sentiment analysis, IEEE Access 8 (2020)
  113878--113899.
\newblock \href {http://dx.doi.org/10.1109/ACCESS.2020.3003625}
  {\path{doi:10.1109/ACCESS.2020.3003625}}.

\bibitem{app11093986}
Z.~Kastrati, F.~Dalipi, A.~S. Imran, K.~Pireva~Nuci, M.~A. Wani, Sentiment
  analysis of students’ feedback with nlp and deep learning: A systematic
  mapping study, Applied Sciences 11~(9).
\newblock \href {http://dx.doi.org/10.3390/app11093986}
  {\path{doi:10.3390/app11093986}}.

\bibitem{Baragash_2021}
R.~Baragash, H.~Aldowah, Sentiment analysis in higher education: a systematic
  mapping review, Journal of Physics: Conference Series 1860~(1) (2021) 012002.
\newblock \href {http://dx.doi.org/10.1088/1742-6596/1860/1/012002}
  {\path{doi:10.1088/1742-6596/1860/1/012002}}.

\bibitem{Ding.2018}
J.~Ding, H.~Sun, X.~Wang, X.~Liu, Entity-level sentiment analysis of issue
  comments, in: 2018 ACM/IEEE 3rd International Workshop on Emotion Awareness
  in Software Engineering, IEEE, Piscataway, NJ, 2018.
\newblock \href {http://dx.doi.org/10.1145/3194932.3194935}
  {\path{doi:10.1145/3194932.3194935}}.

\bibitem{Imtiaz.2018}
N.~Imtiaz, J.~Middleton, P.~Girouard, E.~Murphy-Hill, Sentiment and politeness
  analysis tools on developer discussions are unreliable, but so are people,
  in: Proceedings of the 3rd International Workshop on Emotion Awareness in
  Software Engineering, SEmotion '18, Association for Computing Machinery, New
  York, NY, USA, 2018, p. 55–61.
\newblock \href {http://dx.doi.org/10.1145/3194932.3194938}
  {\path{doi:10.1145/3194932.3194938}}.

\bibitem{StackEmo}
A.~S.~M. Venigalla, S.~Chimalakonda, Stackemo: Towards enhancing user
  experience by augmenting stack overflow with emojis, in: Proceedings of the
  29th ACM Joint Meeting on European Software Engineering Conference and
  Symposium on the Foundations of Software Engineering, ESEC/FSE 2021,
  Association for Computing Machinery, New York, NY, USA, 2021, p. 1550–1554.
\newblock \href {http://dx.doi.org/10.1145/3468264.3473119}
  {\path{doi:10.1145/3468264.3473119}}.

\bibitem{F.Calefato.2015}
F.~Calefato, F.~Lanubile, M.~C. Marasciulo, N.~Novielli, Mining successful
  answers in stack overflow, in: 2015 IEEE/ACM 12th Working Conference on
  Mining Software Repositories, 2015, pp. 430--433.
\newblock \href {http://dx.doi.org/10.1109/MSR.2015.56}
  {\path{doi:10.1109/MSR.2015.56}}.

\bibitem{Umer.2020}
Q.~Umer, H.~Liu, I.~Illahi, Cnn-based automatic prioritization of bug reports,
  IEEE Transactions on Reliability (2020) 1--14\href
  {http://dx.doi.org/10.1109/TR.2019.2959624}
  {\path{doi:10.1109/TR.2019.2959624}}.

\bibitem{Werner.2018}
C.~Werner, G.~Tapuc, L.~Montgomery, D.~Sharma, S.~Dodos, D.~Damian, How angry
  are your customers? sentiment analysis of support tickets that escalate, in:
  D.~Fucci, N.~Novielli, E.~Guzm{\'a}n (Eds.), 2018 1st International Workshop
  on Affective Computing for Requirements Engineering, IEEE, Piscataway, NJ,
  2018.
\newblock \href {http://dx.doi.org/10.1109/affectre.2018.00006}
  {\path{doi:10.1109/affectre.2018.00006}}.

\bibitem{9240704}
T.~Zhang, B.~Xu, F.~Thung, S.~A. Haryono, D.~Lo, L.~Jiang, Sentiment analysis
  for software engineering: How far can pre-trained transformer models go?, in:
  2020 IEEE International Conference on Software Maintenance and Evolution
  (ICSME), 2020, pp. 70--80.
\newblock \href {http://dx.doi.org/10.1109/ICSME46990.2020.00017}
  {\path{doi:10.1109/ICSME46990.2020.00017}}.

\bibitem{Devlin.2019}
J.~Devlin, M.-W. Chang, K.~Lee, K.~Toutanova, Bert: Pre-training of deep
  bidirectional transformers for language understanding (2019).

\bibitem{Novielli.replication2021}
N.~Novielli, F.~Calefato, F.~Lanubile, A.~Serebrenik, Assessment of
  off-the-shelf se-specific sentiment analysis tools: An extended replication
  study, Empirical Software Engineering 26~(4) (2021) 77.
\newblock \href {http://dx.doi.org/10.1007/s10664-021-09960-w}
  {\path{doi:10.1007/s10664-021-09960-w}}.

\bibitem{N.Novielli.2018}
{Nicole Novielli}, {Daniela Girardi}, {Filippo Lanubile}, A benchmark study on
  sentiment analysis for software engineering research, in: 2018 IEEE/ACM 15th
  International Conference on Mining Software Repositories (MSR), 2018, pp.
  364--375.

\bibitem{Biswas.2019}
E.~Biswas, K.~Vijay-Shanker, L.~Pollock, Exploring word embedding techniques to
  improve sentiment analysis of software engineering texts, in: 2019 IEEE/ACM
  16th International Conference on Mining Software Repositories, IEEE,
  Piscataway, NJ, 2019.
\newblock \href {http://dx.doi.org/10.1109/msr.2019.00020}
  {\path{doi:10.1109/msr.2019.00020}}.

\bibitem{Islam.2018c}
M.~R. Islam, M.~F. Zibran, A comparison of software engineering domain specific
  sentiment analysis tools, in: 25th IEEE International Conference on Software
  Analysis, Evolution and Reengineering, IEEE, Piscataway, NJ, 2018.
\newblock \href {http://dx.doi.org/10.1109/saner.2018.8330245}
  {\path{doi:10.1109/saner.2018.8330245}}.

\bibitem{BinLin.2021}
{Bin Lin}, {Nathan W. Cassee}, {Alexander Serebrenik}, {Gabriele Bavota},
  {Nicole Novielli}, {Michele Lanza}, Opinion mining for software development:
  A systematic literature review, ACM Transactions on Software Engineering and
  Methodology XX~(X).

\bibitem{SANCHEZGORDON201923}
M.~Sánchez-Gordón, R.~Colomo-Palacios, Taking the emotional pulse of software
  engineering — a systematic literature review of empirical studies,
  Information and Software Technology 115 (2019) 23--43.
\newblock \href {http://dx.doi.org/10.1016/j.infsof.2019.08.002}
  {\path{doi:10.1016/j.infsof.2019.08.002}}.

\bibitem{Wohlin.2012}
C.~Wohlin, P.~Runeson, M.~Höst, M.~C. Ohlsson, B.~Regnell, A.~Wesslén,
  Experimentation in software engineering, Springer, Berlin, 2012.
\newblock \href {http://dx.doi.org/10.1007/978-3-642-29044-2}
  {\path{doi:10.1007/978-3-642-29044-2}}.

\bibitem{KITCHENHAM20097}
B.~Kitchenham, O.~{Pearl Brereton}, D.~Budgen, M.~Turner, J.~Bailey,
  S.~Linkman, Systematic literature reviews in software engineering – a
  systematic literature review, Information and Software Technology 51~(1)
  (2009) 7 -- 15, special Section - Most Cited Articles in 2002 and Regular
  Research Papers.
\newblock \href {http://dx.doi.org/10.1016/j.infsof.2008.09.009}
  {\path{doi:10.1016/j.infsof.2008.09.009}}.

\bibitem{KitchenhamBA.2007}
{Barbara Kitchenham}, {Stuart M. Charters}, Guidelines for performing
  Systematic Literature Reviews in Software Engineering, Vol.~2, 2007.

\bibitem{Petersen.2008}
K.~Petersen, R.~Feldt, S.~Mujtaba, M.~Mattsson, Systematic mapping studies in
  software engineering, {BCS Learning {\&} Development}, 2008.
\newblock \href {http://dx.doi.org/10.14236/ewic/ease2008.8}
  {\path{doi:10.14236/ewic/ease2008.8}}.

\bibitem{8812836}
C.~Unger-Windeler, J.~Klünder, K.~Schneider, A mapping study on product owners
  in industry: Identifying future research directions, in: 2019 IEEE/ACM
  International Conference on Software and System Processes (ICSSP), 2019, pp.
  135--144.
\newblock \href {http://dx.doi.org/10.1109/ICSSP.2019.00026}
  {\path{doi:10.1109/ICSSP.2019.00026}}.

\bibitem{8984351}
A.~Yasin, R.~Fatima, L.~Wen, W.~Afzal, M.~Azhar, R.~Torkar, On using grey
  literature and google scholar in systematic literature reviews in software
  engineering, IEEE Access 8 (2020) 36226--36243.
\newblock \href {http://dx.doi.org/10.1109/ACCESS.2020.2971712}
  {\path{doi:10.1109/ACCESS.2020.2971712}}.

\bibitem{7929422}
S.~Shevtsov, M.~Berekmeri, D.~Weyns, M.~Maggio, Control-theoretical software
  adaptation: A systematic literature review, IEEE Transactions on Software
  Engineering 44~(8) (2018) 784--810.
\newblock \href {http://dx.doi.org/10.1109/TSE.2017.2704579}
  {\path{doi:10.1109/TSE.2017.2704579}}.

\bibitem{kosa.2016}
M.~Kosa, M.~Yilmaz, R.~O’Connor, P.~Clarke, Software engineering education
  and games: A systematic literature review, Journal of Universal Computer
  Science 22 (2016) 1558–1574.

\bibitem{GAROUSI2016106}
V.~Garousi, K.~Petersen, B.~Ozkan, Challenges and best practices in
  industry-academia collaborations in software engineering: A systematic
  literature review, Information and Software Technology 79 (2016) 106--127.
\newblock \href {http://dx.doi.org/10.1016/j.infsof.2016.07.006}
  {\path{doi:10.1016/j.infsof.2016.07.006}}.

\bibitem{Klunder.2019}
J.~A.-C. Kl{\"u}nder, P.~Hohl, N.~Prenner, K.~Schneider, Transformation towards
  agile software product line engineering in large companies: A literature
  review, Journal of Software: Evolution and Process 31~(5) (2019) e2168.
\newblock \href {http://dx.doi.org/10.1002/smr.2168}
  {\path{doi:10.1002/smr.2168}}.

\bibitem{10.1145/3029387.3029392}
K.~Zahra, F.~Azam, F.~Ilyas, H.~Faisal, N.~Ambreen, N.~Gondal, Success factors
  of organizational change in software process improvement: A systematic
  literature review, in: Proceedings of the 5th International Conference on
  Information and Education Technology, ICIET '17, Association for Computing
  Machinery, New York, NY, USA, 2017, p. 155–160.
\newblock \href {http://dx.doi.org/10.1145/3029387.3029392}
  {\path{doi:10.1145/3029387.3029392}}.

\bibitem{10.1145/3379177.3388907}
N.~Prenner, C.~Unger-Windeler, K.~Schneider, How are hybrid development
  approaches organized? a systematic literature review, in: Proceedings of the
  International Conference on Software and System Processes, ICSSP '20,
  Association for Computing Machinery, New York, NY, USA, 2020, p. 145–154.
\newblock \href {http://dx.doi.org/10.1145/3379177.3388907}
  {\path{doi:10.1145/3379177.3388907}}.

\bibitem{8389299}
M.~Deshpande, V.~Rao, Depression detection using emotion artificial
  intelligence, in: 2017 International Conference on Intelligent Sustainable
  Systems (ICISS), 2017, pp. 858--862.
\newblock \href {http://dx.doi.org/10.1109/ISS1.2017.8389299}
  {\path{doi:10.1109/ISS1.2017.8389299}}.

\bibitem{Liu.2012}
B.~Liu, Sentiment analysis and opinion mining, Synthesis Lectures on Human
  Language Technologies 5~(1) (2012) 1--167.
\newblock \href {http://dx.doi.org/10.2200/S00416ED1V01Y201204HLT016}
  {\path{doi:10.2200/S00416ED1V01Y201204HLT016}}.

\bibitem{Liu2012}
B.~Liu, L.~Zhang, A Survey of Opinion Mining and Sentiment Analysis, Springer
  US, Boston, MA, 2012, pp. 415--463.
\newblock \href {http://dx.doi.org/10.1007/978-1-4614-3223-4_13}
  {\path{doi:10.1007/978-1-4614-3223-4_13}}.

\bibitem{Fang.2015}
X.~Fang, J.~Zhan, Sentiment analysis using product review data, Journal of Big
  Data 2~(1) (2015) 1--14.
\newblock \href {http://dx.doi.org/10.1186/s40537-015-0015-2}
  {\path{doi:10.1186/s40537-015-0015-2}}.

\bibitem{Ortu.2016}
M.~Ortu, A.~Murgia, G.~Destefanis, P.~Tourani, R.~Tonelli, M.~Marchesi,
  B.~Adams, The emotional side of software developers in jira, in: Proceedings
  of the 13th International Conference on Mining Software Repositories, MSR
  '16, Association for Computing Machinery, New York, NY, USA, 2016, p.
  480–483.
\newblock \href {http://dx.doi.org/10.1145/2901739.2903505}
  {\path{doi:10.1145/2901739.2903505}}.

\bibitem{parrott2001emotions}
W.~G. Parrott, Emotions in social psychology: Essential readings, psychology
  press, 2001.

\bibitem{Wohlin.2014}
C.~Wohlin, Guidelines for snowballing in systematic literature studies and a
  replication in software engineering, in: Proceedings of the 18th
  International Conference on Evaluation and Assessment in Software
  Engineering, EASE '14, Association for Computing Machinery, New York, NY,
  USA, 2014.
\newblock \href {http://dx.doi.org/10.1145/2601248.2601268}
  {\path{doi:10.1145/2601248.2601268}}.

\bibitem{martin_obaidi_2021_4726650}
M.~Obaidi, L.~Nagel, A.~Specht, J.~Klünder, {Dataset: Systematic Mapping Study
  on the Development and Application of Sentiment Analysis Tools in Software
  Engineering} (Mar. 2022).
\newblock \href {http://dx.doi.org/10.5281/zenodo.4726650}
  {\path{doi:10.5281/zenodo.4726650}}.

\bibitem{Islam.2016}
M.~R. Islam, M.~F. Zibran, Towards understanding and exploiting developers'
  emotional variations in software engineering, in: Y.-T. Song (Ed.), 2016
  IEEE/ACIS 14th International Conference on Software Engineering Research,
  Management and Applications (SERA), IEEE, Piscataway, NJ, 2016, pp. 185--192.
\newblock \href {http://dx.doi.org/10.1109/SERA.2016.7516145}
  {\path{doi:10.1109/SERA.2016.7516145}}.

\bibitem{10.1145/3424308}
Z.~Chen, Y.~Cao, H.~Yao, X.~Lu, X.~Peng, H.~Mei, X.~Liu, Emoji-powered
  sentiment and emotion detection from software developers’ communication
  data, ACM Trans. Softw. Eng. Methodol. 30~(2).
\newblock \href {http://dx.doi.org/10.1145/3424308}
  {\path{doi:10.1145/3424308}}.

\bibitem{9492202}
J.~Wu, C.~Ye, H.~Zhou, Bert for sentiment classification in software
  engineering, in: 2021 International Conference on Service Science (ICSS),
  2021, pp. 115--121.
\newblock \href {http://dx.doi.org/10.1109/ICSS53362.2021.00026}
  {\path{doi:10.1109/ICSS53362.2021.00026}}.

\bibitem{Calefato.2019}
F.~Calefato, F.~Lanubile, N.~Novielli, L.~Quaranta, Emtk - the emotion mining
  toolkit, in: 2019 IEEE/ACM 4th International Workshop on Emotion Awareness in
  Software Engineering (SEmotion 2019), IEEE, Piscataway, NJ, 2019, pp. 34--37.
\newblock \href {http://dx.doi.org/10.1109/SEmotion.2019.00014}
  {\path{doi:10.1109/SEmotion.2019.00014}}.

\bibitem{10.1145/3392877}
M.~E. Whiting, I.~Gao, M.~Xing, N.~J. Diarrassouba, T.~Nguyen, M.~S. Bernstein,
  Parallel worlds: Repeated initializations of the same team to improve team
  viability 4~(CSCW1).
\newblock \href {http://dx.doi.org/10.1145/3392877}
  {\path{doi:10.1145/3392877}}.

\bibitem{Gkontzis.2017}
A.~F. Gkontzis, C.~V. Karachristos, C.~T. Panagiotakopoulos, E.~C.
  Stavropoulos, V.~S. Verykios, Sentiment analysis to track emotion and
  polarity in student fora, in: V.~Vlachos (Ed.), Proceedings of the 21st
  Pan-Hellenic Conference on Informatics, ACM, New York, NY, 2017.
\newblock \href {http://dx.doi.org/10.1145/3139367.3139389}
  {\path{doi:10.1145/3139367.3139389}}.

\bibitem{Guzman.2013b}
E.~Guzman, Visualizing emotions in software development projects, in: A.~Telea
  (Ed.), 2013 First IEEE Working Conference on Software Visualization
  (VISSOFT), IEEE, Piscataway, NJ, 2013.
\newblock \href {http://dx.doi.org/10.1109/vissoft.2013.6650529}
  {\path{doi:10.1109/vissoft.2013.6650529}}.

\bibitem{Guzman.2013}
E.~Guzman, B.~Bruegge, Towards emotional awareness in software development
  teams, in: B.~Meyer, M.~Mezini, L.~Baresi (Eds.), 2013 9th Joint Meeting of
  the European Software Engineering Conference and the ACM SIGSOFT Symposium on
  the Foundations of Software Engineering (ESEC/FSE) : proceedings : August
  18-26, 2013, Saint Petersburg, Russia, ESEC/FSE 2013, ACM, 2013, p.
  671–674.
\newblock \href {http://dx.doi.org/10.1145/2491411.2494578}
  {\path{doi:10.1145/2491411.2494578}}.

\bibitem{ElHalees.2014}
A.~M. El-Halees, Software usability evaluation using opinion mining, Journal of
  Software 9~(2).
\newblock \href {http://dx.doi.org/10.4304/jsw.9.2.343-349}
  {\path{doi:10.4304/jsw.9.2.343-349}}.

\bibitem{Nayebi.2017}
M.~Nayebi, H.~Farahi, G.~Ruhe, Which version should be released to app store?,
  in: 11th ACM/IEEE International Symposium on Empirical Software Engineering
  and Measurement, IEEE, Piscataway, NJ, 2017, pp. 324--333.
\newblock \href {http://dx.doi.org/10.1109/ESEM.2017.46}
  {\path{doi:10.1109/ESEM.2017.46}}.

\bibitem{.2015}
P.~Kaewyong, A.~Sukprasert, N.~Salim, F.~Phang, The possibility of students'
  comments automatic interpret using lexicon based sentiment analysis to
  teacher evaluation, 2015.

\bibitem{Aung.2017}
K.~Z. Aung, N.~N. Myo, Sentiment analysis of students' comment using lexicon
  based approach, in: G.~Zhu (Ed.), 16th IEEE/ACIS International Conference on
  Computer and Information Science (ICIS 2017), IEEE, Piscataway, NJ, 2017, pp.
  149--154.
\newblock \href {http://dx.doi.org/10.1109/icis.2017.7959985}
  {\path{doi:10.1109/icis.2017.7959985}}.

\bibitem{Thelwall.2010}
M.~Thelwall, K.~Buckley, G.~Paltoglou, {Di Cai}, A.~Kappas, Sentiment strength
  detection in short informal text, Journal of the American Society for
  Information Science and Technology 61~(12) (2010) 2544--2558.
\newblock \href {http://dx.doi.org/10.1002/asi.21416}
  {\path{doi:10.1002/asi.21416}}.

\bibitem{Thelwall.2012}
M.~Thelwall, K.~Buckley, G.~Paltoglou, Sentiment strength detection for the
  social web, Journal of the American Society for Information Science and
  Technology 63~(1) (2012) 163--173.
\newblock \href {http://dx.doi.org/10.1002/asi.21662}
  {\path{doi:10.1002/asi.21662}}.

\bibitem{Cagnoni.2020}
S.~Cagnoni, L.~Cozzini, G.~Lombardo, M.~Mordonini, A.~Poggi, M.~Tomaiuolo,
  Emotion-based analysis of programming languages on stack overflow, ICT
  Express 6~(3) (2020) 238--242.
\newblock \href {http://dx.doi.org/10.1016/j.icte.2020.07.002}
  {\path{doi:10.1016/j.icte.2020.07.002}}.

\bibitem{Lin.2019}
B.~Lin, F.~Zampetti, G.~Bavota, M.~{Di Penta}, M.~Lanza, Pattern-based mining
  of opinions in q{\&}a websites, in: 2019 IEEE/ACM 41st International
  Conference on Software Engineering, IEEE, Piscataway, NJ, 2019.
\newblock \href {http://dx.doi.org/10.1109/icse.2019.00066}
  {\path{doi:10.1109/icse.2019.00066}}.

\bibitem{10.1007/978-3-030-64266-2_8}
J.~Kl{\"u}nder, J.~Horstmann, O.~Karras, Identifying the mood of a software
  development team by analyzing text-based communication in chats with machine
  learning, in: R.~Bernhaupt, C.~Ardito, S.~Sauer (Eds.), Human-Centered
  Software Engineering, Springer International Publishing, Cham, 2020, pp.
  133--151.

\bibitem{10.1145/3442442.3458612}
T.~Yang, C.~Gao, J.~Zang, D.~Lo, M.~Lyu, Tour: Dynamic topic and sentiment
  analysis of user reviews for assisting app release, in: Companion Proceedings
  of the Web Conference 2021, WWW '21, Association for Computing Machinery, New
  York, NY, USA, 2021, p. 708–712.
\newblock \href {http://dx.doi.org/10.1145/3442442.3458612}
  {\path{doi:10.1145/3442442.3458612}}.

\bibitem{kumar2017opinion}
A.~Kumar, A.~Abraham, Opinion mining to assist user acceptance testing for
  open-beta versions., Journal of Information Assurance \& Security 12~(4).

\bibitem{Islam.2019}
M.~R. Islam, M.~K. Ahmmed, M.~F. Zibran, Marvalous - machine learning based
  detection of emotions in the valence-arousal space in software engineering
  text, in: C.-C. Hung (Ed.), Proceedings of the 34th ACMSIGAPP Symposium on
  Applied Computing, ACM Digital Library, ACM, New York, NY, 2019, pp.
  1786--1793.
\newblock \href {http://dx.doi.org/10.1145/3297280.3297455}
  {\path{doi:10.1145/3297280.3297455}}.

\bibitem{8890439}
J.~Shen, O.~Baysal, M.~O. Shafiq, Evaluating the performance of machine
  learning sentiment analysis algorithms in software engineering, in: 2019 IEEE
  Intl Conf on Dependable, Autonomic and Secure Computing, Intl Conf on
  Pervasive Intelligence and Computing, Intl Conf on Cloud and Big Data
  Computing, Intl Conf on Cyber Science and Technology Congress
  (DASC/PiCom/CBDCom/CyberSciTech), 2019, pp. 1023--1030.
\newblock \href
  {http://dx.doi.org/10.1109/DASC/PiCom/CBDCom/CyberSciTech.2019.00185}
  {\path{doi:10.1109/DASC/PiCom/CBDCom/CyberSciTech.2019.00185}}.

\bibitem{mostafa2018investigating}
L.~Mostafa, M.~Abd~Elghany, Investigating game developers’ guilt emotions
  using sentiment analysis, Int. J. Softw. Eng. Appl.(IJSEA) 9 (2018) 16.
\newblock \href {http://dx.doi.org/10.5121/ijsea.2018.9604}
  {\path{doi:10.5121/ijsea.2018.9604}}.

\bibitem{Murgia.2018}
A.~Murgia, M.~Ortu, P.~Tourani, B.~Adams, S.~Demeyer, An exploratory
  qualitative and quantitative analysis of emotions in issue report comments of
  open source systems, Empirical Software Engineering 23~(1) (2018) 521--564.
\newblock \href {http://dx.doi.org/10.1007/s10664-017-9526-0}
  {\path{doi:10.1007/s10664-017-9526-0}}.

\bibitem{Patwardhan.2017}
A.~Patwardhan, Sentiment identification for collaborative, geographically
  dispersed, cross-functional software development teams, in: 2017 IEEE 3rd
  International Conference on Collaboration and Internet Computing, IEEE,
  Piscataway, NJ, 2017.
\newblock \href {http://dx.doi.org/10.1109/cic.2017.00014}
  {\path{doi:10.1109/cic.2017.00014}}.

\bibitem{emotionlevels}
P.~Shaver, J.~Schwartz, D.~Kirson, C.~O'Connor, Emotion knowledge: Further
  exploration of a prototype approach, Journal of Personality and Social
  Psychology 52~(6) (1987) 1061--1086.
\newblock \href {http://dx.doi.org/10.1037/0022-3514.52.6.1061}
  {\path{doi:10.1037/0022-3514.52.6.1061}}.

\bibitem{10.1145/3463274.3463805}
J.~Cheriyan, B.~T.~R. Savarimuthu, S.~Cranefield, Towards offensive language
  detection and reduction in four software engineering communities, EASE 2021,
  Association for Computing Machinery, New York, NY, USA, 2021, p. 254–259.
\newblock \href {http://dx.doi.org/10.1145/3463274.3463805}
  {\path{doi:10.1145/3463274.3463805}}.

\bibitem{Loper.17.05.2002}
E.~Loper, S.~Bird, Nltk: The natural language toolkit (2002).

\bibitem{liu2019roberta}
Y.~Liu, M.~Ott, N.~Goyal, J.~Du, M.~Joshi, D.~Chen, O.~Levy, M.~Lewis,
  L.~Zettlemoyer, V.~Stoyanov, Roberta: A robustly optimized bert pretraining
  approach (2019).

\bibitem{396988}
G.~Holmes, A.~Donkin, I.~Witten, Weka: a machine learning workbench, in:
  Proceedings of ANZIIS '94 - Australian New Zealnd Intelligent Information
  Systems Conference, 1994, pp. 357--361.
\newblock \href {http://dx.doi.org/10.1109/ANZIIS.1994.396988}
  {\path{doi:10.1109/ANZIIS.1994.396988}}.

\bibitem{lan2020albert}
Z.~Lan, M.~Chen, S.~Goodman, K.~Gimpel, P.~Sharma, R.~Soricut, Albert: A lite
  bert for self-supervised learning of language representations (2020).
\newblock \href {http://arxiv.org/abs/1909.11942} {\path{arXiv:1909.11942}}.

\bibitem{NEURIPS2019_dc6a7e65}
Z.~Yang, Z.~Dai, Y.~Yang, J.~Carbonell, R.~R. Salakhutdinov, Q.~V. Le, Xlnet:
  Generalized autoregressive pretraining for language understanding, in:
  H.~Wallach, H.~Larochelle, A.~Beygelzimer, F.~d\textquotesingle
  Alch\'{e}-Buc, E.~Fox, R.~Garnett (Eds.), Advances in Neural Information
  Processing Systems, Vol.~32, Curran Associates, Inc., 2019.

\bibitem{Manning.2014}
C.~Manning, M.~Surdeanu, J.~Bauer, J.~Finkel, S.~Bethard, D.~McClosky, The
  stanford corenlp natural language processing toolkit, in: Proceedings of 52nd
  Annual Meeting of the Association for Computational Linguistics: System
  Demonstrations, {Association for Computational Linguistics}, Stroudsburg, PA,
  USA, 2014.
\newblock \href {http://dx.doi.org/10.3115/v1/p14-5010}
  {\path{doi:10.3115/v1/p14-5010}}.

\bibitem{Jongeling.2017}
R.~Jongeling, P.~Sarkar, S.~Datta, A.~Serebrenik, On negative results when
  using sentiment analysis tools for software engineering research, Empirical
  Software Engineering 22~(5) (2017) 2543--2584.
\newblock \href {http://dx.doi.org/10.1007/s10664-016-9493-x}
  {\path{doi:10.1007/s10664-016-9493-x}}.

\bibitem{Kaur.2018}
A.~Kaur, A.~P. Singh, G.~S. Dhillon, D.~Bisht, Emotion mining and sentiment
  analysis in software engineering domain, in: Proceedings of the Second
  International Conference on Electronics, Communication and Aerospace
  Technology (ICECA 2018), IEEE, Piscataway, NJ, 2018, pp. 1170--1173.
\newblock \href {http://dx.doi.org/10.1109/ICECA.2018.8474619}
  {\path{doi:10.1109/ICECA.2018.8474619}}.

\bibitem{10.1145/3479497}
I.~Ferreira, J.~Cheng, B.~Adams, The "shut the f**k up" phenomenon:
  Characterizing incivility in open source code review discussions 5~(CSCW2).
\newblock \href {http://dx.doi.org/10.1145/3479497}
  {\path{doi:10.1145/3479497}}.

\bibitem{8643972}
G.~Uddin, F.~Khomh, Automatic mining of opinions expressed about apis in stack
  overflow, IEEE Transactions on Software Engineering 47~(3) (2021) 522--559.
\newblock \href {http://dx.doi.org/10.1109/TSE.2019.2900245}
  {\path{doi:10.1109/TSE.2019.2900245}}.

\bibitem{CABRERADIEGO2020105633}
L.~A. Cabrera-Diego, N.~Bessis, I.~Korkontzelos, Classifying emotions in stack
  overflow and jira using a multi-label approach, Knowledge-Based Systems 195
  (2020) 105633.
\newblock \href {http://dx.doi.org/10.1016/j.knosys.2020.105633}
  {\path{doi:10.1016/j.knosys.2020.105633}}.

\bibitem{9474673}
K.-i. Park, B.~Sharif, Assessing perceived sentiment in pull requests with
  emoji: Evidence from tools and developer eye movements, in: 2021 IEEE/ACM
  Sixth International Workshop on Emotion Awareness in Software Engineering
  (SEmotion), 2021, pp. 1--6.
\newblock \href {http://dx.doi.org/10.1109/SEmotion52567.2021.00009}
  {\path{doi:10.1109/SEmotion52567.2021.00009}}.

\bibitem{.2020}
N.~Novielli, F.~Calefato, D.~Dongiovanni, D.~Girardi, F.~Lanubile, A gold
  standard for polarity of emotions of software developers in github (2020).
\newblock \href {http://dx.doi.org/10.6084/m9.figshare.11604597}
  {\path{doi:10.6084/m9.figshare.11604597}}.

\bibitem{.2018}
M.~R. Islam, M.~F. Zibran, Sentiment Analysis of Software Bug Related Commit
  Messages, ISCA, 2018.

\bibitem{10.1145/3180155.3180195}
B.~Lin, F.~Zampetti, G.~Bavota, M.~Di~Penta, M.~Lanza, R.~Oliveto, Sentiment
  analysis for software engineering: How far can we go?, in: Proceedings of the
  40th International Conference on Software Engineering, ICSE '18, Association
  for Computing Machinery, New York, NY, USA, 2018, p. 94–104.
\newblock \href {http://dx.doi.org/10.1145/3180155.3180195}
  {\path{doi:10.1145/3180155.3180195}}.

\bibitem{9462971}
K.~Sun, H.~Gao, H.~Kuang, X.~Ma, G.~Rong, D.~Shao, H.~Zhang, Exploiting the
  unique expression for improved sentiment analysis in software engineering
  text, in: 2021 IEEE/ACM 29th International Conference on Program
  Comprehension (ICPC), 2021, pp. 149--159.
\newblock \href {http://dx.doi.org/10.1109/ICPC52881.2021.00023}
  {\path{doi:10.1109/ICPC52881.2021.00023}}.

\bibitem{Fabry.2021}
R.~E. Fabry, Getting it: A predictive processing approach to irony
  comprehension, Synthese 198~(7) (2021) 6455--6489.
\newblock \href {http://dx.doi.org/10.1007/s11229-019-02470-9}
  {\path{doi:10.1007/s11229-019-02470-9}}.

\bibitem{e23040394}
R.~Akula, I.~Garibay, Interpretable multi-head self-attention architecture for
  sarcasm detection in social media, Entropy 23~(4).
\newblock \href {http://dx.doi.org/10.3390/e23040394}
  {\path{doi:10.3390/e23040394}}.

\bibitem{10.5555/2735522.2735528}
P.~Tourani, Y.~Jiang, B.~Adams, Monitoring sentiment in open source mailing
  lists: Exploratory study on the apache ecosystem, in: Proceedings of 24th
  Annual International Conference on Computer Science and Software Engineering,
  CASCON '14, IBM Corp., USA, 2014, p. 34–44.

\bibitem{Ahasanuzzaman.2020}
M.~Ahasanuzzaman, M.~Asaduzzaman, C.~K. Roy, K.~A. Schneider, Caps: a
  supervised technique for classifying stack overflow posts concerning api
  issues, Empirical Software Engineering 25~(2) (2020) 1493--1532.
\newblock \href {http://dx.doi.org/10.1007/s10664-019-09743-4}
  {\path{doi:10.1007/s10664-019-09743-4}}.

\bibitem{Guzman.2017}
E.~Guzman, R.~Alkadhi, N.~Seyff, An exploratory study of twitter messages about
  software applications, Requirements Engineering 22~(3) (2017) 387--412.
\newblock \href {http://dx.doi.org/10.1007/s00766-017-0274-x}
  {\path{doi:10.1007/s00766-017-0274-x}}.

\bibitem{Novielli.2015}
N.~Novielli, F.~Calefato, F.~Lanubile, The challenges of sentiment detection in
  the social programmer ecosystem, in: I.~Hammouda, A.~Sillitti (Eds.),
  Proceedings of the 7th International Workshop on Social Software Engineering
  - SSE 2015, SSE 2015, {ACM Press}, New York, New York, USA, 2015, pp. 33--40.
\newblock \href {http://dx.doi.org/10.1145/2804381.2804387}
  {\path{doi:10.1145/2804381.2804387}}.

\bibitem{S.Panichella.2015}
S.~Panichella, A.~Di~Sorbo, E.~Guzman, C.~A. Visaggio, G.~Canfora, H.~C. Gall,
  How can i improve my app? classifying user reviews for software maintenance
  and evolution, in: 2015 IEEE International Conference on Software Maintenance
  and Evolution (ICSME), 2015, pp. 281--290.
\newblock \href {http://dx.doi.org/10.1109/ICSM.2015.7332474}
  {\path{doi:10.1109/ICSM.2015.7332474}}.

\bibitem{9474682}
N.~Mansoor, C.~S. Peterson, B.~Sharif, How developers and tools categorize
  sentiment in stack overflow questions - a pilot study, in: 2021 IEEE/ACM
  Sixth International Workshop on Emotion Awareness in Software Engineering
  (SEmotion), 2021, pp. 19--22.
\newblock \href {http://dx.doi.org/10.1109/SEmotion52567.2021.00012}
  {\path{doi:10.1109/SEmotion52567.2021.00012}}.

\bibitem{PLUTCHIK19803}
R.~Plutchik, Chapter 1 - a general psychoevolutionary theory of emotion, in:
  R.~Plutchik, H.~Kellerman (Eds.), Theories of Emotion, Academic Press, 1980,
  pp. 3--33.
\newblock \href {http://dx.doi.org/10.1016/B978-0-12-558701-3.50007-7}
  {\path{doi:10.1016/B978-0-12-558701-3.50007-7}}.

\bibitem{Watson.1988}
D.~Watson, L.~Clark, A.~Tellegen, Development and validation of brief measures
  of positive and negative affect: the panas scales, Journal of personality and
  social psychology 54~(6) (1988) 1063--1070.
\newblock \href {http://dx.doi.org/10.1037//0022-3514.54.6.1063}
  {\path{doi:10.1037//0022-3514.54.6.1063}}.

\bibitem{asyrofi2021biasfinder}
M.~H. Asyrofi, Z.~Yang, I.~N.~B. Yusuf, H.~J. Kang, F.~Thung, D.~Lo,
  Biasfinder: Metamorphic test generation to uncover bias for sentiment
  analysis systems (2021).
\newblock \href {http://arxiv.org/abs/2102.01859} {\path{arXiv:2102.01859}}.

\end{thebibliography}

\end{document}